\documentstyle[aps,preprint,epsfig]{revtex}

\begin{document}
\draft
\title{Complexity, multiresolution, non-stationarity and entropic scaling:
Teen birth thermodynamics}
\author{Nicola Scafetta$^{1}$ and
Bruce J. West$^{1,2}$.}
\address{$^1$ Department of Physics, FELL,  Duke University, P.O. Box 90319, Durham, North Carolina
27708. }
\address{$^{2}$ Mathematics Division, Army Research Office, Research Triangle Park, NC
27709. }
\date{\today}
\maketitle

\newpage

\begin{abstract}
This paper presents a statistical methodology for analyzing a complex
phenomenon in which deterministic and scaling components are superimposed.
Our approach is based on the wavelet multiresolution analysis combined with
the scaling analysis of the entropy of a time series. The wavelet
multiresolution analysis decomposes the signal in a {\it scale-by-scale}
manner. The {\it scale-by-scale} decomposition generates smooth and detail
curves which are evaluated and studied. A wavelet-based smoothing filtering
is used to estimate the daily birth rate and conception rate during the
year. The scaling analysis is based upon the Diffusion Entropy Analysis
(DEA). The joint use of the DEA and the wavelet multiresolution analysis
allows: 1) the separation of the deterministic and, therefore, non-scaling
component from the scaling component of the signal; 2) the determination of
the stochastic information characterizing the teen birth phenomenon at each
time scale. The daily data cover the number of births to teens in Texas
during the period from 1964 to 1999.\\

{\bf {\it KEYWORDS}: fractals, time series, wavelet, teenagers, births   }
\end{abstract}




\newpage

\section{Introduction}

No physical measurement is absolutely reproducible, so experiments yield
results in the form of apparently random time series, $\{\xi _{i}\}$, for
physical observables. Typically these time series contain both a slow and a
fast variation. In the engineering literature the slow regular variation $%
f(i)$ of a time series is referred to as the {\it signal} and the rapid
erratic fluctuations $\{\eta _{i}\}$ are called the {\it noise} according to
the equation
\begin{equation}
\xi _{i}=f(i)+\eta _{i}~.  \label{noiff}
\end{equation}
However, the low-frequency, slowly varying part of the spectrum may be
coupled to, and exchanges energy with, the high-frequency, rapidly varying
part of the spectrum; a fact that may result in fractal correlations. Thus,
the traditional view of a deterministic, predictable signal given by the low
frequency or smooth part of the time series on which random, unpredictable
noise is superposed, is usually inappropriate for the study of complex
phenomena. For our study of time series we require a technique that is able
to isolate and separate the deterministic part from the scaling part of the
signal, without distorting the mutual influence of the low-frequency and
high-frequency components of the same time series. Such a technique is
introduced below.

The importance of searching for stationary properties in time series has
become most evident in the areas of biophysics, econophysics and
sociophysics, where the apparent non-stationarity may only, in part, be
associated with simple deterministic mechanisms like periodicities and
linear trends. In the case of the teen birth phenomenon studied in this
paper, many simple sources of evident non-stationary deterministic
components, ranging from weekly to annual periodicity, are observed. It is
worth stressing that these periodicities are perceived as a deviation from
the stationary condition when the statistical analysis is made on a time
scale comparable with the corresponding time period\cite
{solarflares,thermodynamicofsocialprocesses}. One of the main goals of this
paper is to assess the influence of yearly periodicities, so we devote
special attention to detrending annual cycles and to discussing the
sociological meaning of the residual memory after this kind of detrending is
performed.

Our method of analysis is based on the joint use of wavelet analysis and the
Diffusion Entropy Analysis (DEA). The DEA was born with the analysis of the
teen birth phenomenon \cite{thermodynamicofsocialprocesses}, and rapidly
grew up as a very efficient method to detect the scaling of a complex
process. Such complex processes are defined through the scaling of the
probability density function (pdf) of the diffusion process generated by
time series imagined as a physical source of fluctuations. For a detailed
account of this technique as an efficient method of scaling detection the
interested reader can consult Ref. \cite{scalingdetection} and for its
connection with the compression algorithm, see Ref. \cite
{compressionalgorithm}. The wavelets are a powerful method of analysis\cite
{percival} that localizes a signal simultaneously in both time and
frequency. This paper aims at showing the advantages stemming from the joint
use of DEA and wavelets on sociological data sets. We can say that the
adoption of wavelets serves not only the purpose of decomposing the original
signal on a {\it scale-by-scale} basis, but also of detrending from the data
the deterministic component associated with cycles. In this way the relative
rate of the annual cycle of births and conceptions is evaluated, studied and
detrended from the data. Finally, the adoption of the DEA reveals the
scaling of the data and sheds light on the statistics of the fluctuations
around the non-stationary bias.

The outline of the paper is as follows. In Sec. 2 we introduce the basic
analysis of the data of births to teenagers in Texas. In Sec. 3, to make the
paper as self-contained as possible, we present a short review of the
wavelet multiresolution analysis. Sec. 4 is devoted to the study of births
to teenagers based on the wavelet multiresolution analysis. Sec. 5 contains
a short review of the DEA and its application to the data. Sec. 6 is devoted
to a physical model that explains the results of the previous section. Sec.
7 shows the benefits of the joint use of wavelets and DEA for a {\it %
scale-by-scale} analysis of the data. Finally, in Sec. 8 we draw some
conclusion.

Finally, readers familiar with the field of sociology might be confused by
the term {\em rate} that we use in this paper. To avoid confusion, it is
worth remarking that we are not referring here to the epidemiologic use of
the term rate, which is the frequency of a demographic event in a specified
period of time divided by the population at risk \cite{texas}. We use the
term rate in the conventional sense of statistical mechanics. This means
that birth and conception rates correspond to the number of births and
conceptions per unit of time, the unit of time, in our case, being either
one day or (in one case only) one year. 

\section{Births to teenagers in Texas}

Here we introduce the readers to the data that is analyzed in this paper.
The data is the daily count of every birth to a mother under the age of 20
in Texas for 36 years, from 1964 through 1999. Data were obtained from the
Bureau of Vital Statistics in the Texas Department of Health \cite{texas}
and represent all birth records available electronically at the time of the
study. The interested reader can find additional information in Ref. \cite
{thermodynamicofsocialprocesses} and Ref. \cite{morememory} that uses an
approach based upon the relative dispersion. Fig. 1 shows the daily number
of births to teenagers in Texas during the period 1964 through 1999 and
contains 13149 data points. The data show an evident annual cycle that
fluctuates around a complex bias that may be due to social changes during
these years. It is difficult to see in the figure, but more detailed
analysis shows the presence of a strong weekly periodicity due to the
separation of the workweek into weekdays and weekends. The average birth
rate is 132 births per day; the maximum is 226 births per day and the
minimum is 73 births per day; the standard deviation is 21 births per day;
the Skewness is 0.41; the Kurtosis is 0.16.

By fitting the data shown in Fig. 1 with the straight line $y = a+b(t-1964)$%
, where $y$ and $t$ refer to the ordinate and the abscissa, respectively,
using the year as a time unit, we find $a=114$ and $b=0.98$. This means that
the average number of births increases by almost one unit each year. At the
beginning of 1964 the average rate is 114 births per day, and at the end of
1999 the average rate is 150 births per day. Fig. 1 shows also an increasing
range of the number of births during the period 1990-1999.

Fig. 2 shows the spectrum of the data as a function of the time period (the
inverse of frequency) measured in days. The data are previously detrended of
the linear ramp $a+b(x-1964)$ discussed above because Fourier theorem
requires periodic condition at the borders and an artificial discontinuity
at the borders may distort the frequencies. As anticipated, there are two
main periodicities: the annual (365 days) cycle related to the seasonal
change of the light and temperature that affect the human fertility \cite
{lammiron} and the social behavior of the people; the weekly (7 days) cycle
due to workdays and weekends. There are also three other periodicities.
These are: a periodicity of one half year that, for example, may be related
to the autumn and spring school semesters and two weaker periodicities of 73
days and 3.5 days, respectively. All these additional contributions
establish that the annual cycle is not rigorously harmonic even though in
Section 6 we shall use a sinusoidal function to mimic the periodicity effect.

The figure shows also that the background noise exhibits a complex behavior.
For a period shorter than one or two months the spectrum is almost flat, a
typical characteristic of the white or random noise. For larger periods the
background noise shows a scaling behavior similar to that of a fractional
Gaussian noise (FGN) that seems to last for many years. We recall that FGN
is a particular type of correlated noise defined by a spectrum satisfying
the inverse power law
\begin{equation}
P(f)\propto f^{-\beta }~,  \label{fbnps}
\end{equation}
where $\beta =(2H-1)$ and $H$ is the Hurst coefficient. The curve (\ref
{fbnps}), when plotted on log-log graph paper against the period $\tau =1/f$%
, is a straight line with a slope given by $\beta $.

This paper addresses several questions concerning these data. First, by
using a {\it scale-by-scale} definition of ``noise", based upon the wavelet
details, we determine how to eliminate the noise in such a way as to extract
information at each wavelet temporal scale from the data. Second, we discuss
the origin of the increase of the spreading of the number of births per day
during the years 1990-1999 to determine if its origin based upon an annual,
monthly or weekly change. Third, we determine how to estimate the relative
birth and conception rates during the year. Fourth, we discuss the
stochastic properties of the background noise by separating it from the
deterministic annual cycle of the signal. The annual and half annual
periodicity are determined by two wavelet detail curves.


\section{Wavelet multiresolution analysis}

Wavelet analysis \cite{percival} is a powerful method to analyze
time series and is attracting the attention of an ever increasing
number of investigators. Wavelet Transforms makes use of scaling
functions, the wavelets, with the important property of being
localized in both time and frequency. 
A scaling coefficient $\tau$ characterizes the width of a wavelet
function.

Given a signal $x(u)$, the Continuous Wavelet Transform is defined by
\begin{equation}  \label{cwtdhjj3}
W(\tau, t)=\int\limits_{-\infty }^{\infty } \tilde{\psi}_{\tau
,t}(u)~x(u)~du~,
\end{equation}
where the kernel $\tilde{\psi}_{\tau ,t}(u)$ is the wavelet filter. The
original signal can be recovered from its continuous wavelet trasform (CWT)
via the inverse transformation
\begin{equation}  \label{invcwtxu9}
x(u)=\frac{1}{C_{\tilde{\psi}}} \int\limits_{0}^{\infty} \left[
\int\limits_{-\infty }^{\infty } W(\tau,t) ~\tilde{\psi}_{\tau,t}(u)~dt
\right] ~\frac{d\tau}{\tau^2}~,
\end{equation}
where $C_{\tilde{\psi}}$ is a constant related to the Fourier
Transform of the wavelet, see Ref. \cite{percival} for details.
The double integral of Eq. (\ref{invcwtxu9}) suggests that the
original signal may be decomposed into ``continuous details'' that
depend on the scale width $\tau$. However, in the present context,
CWT is not convenient because of the discrete nature of our time
series. There exists a discrete version of the wavelet transform,
the Maximal Overlap Discrete Wavelet Transform (MODWT) \cite
{percival} that is almost independent of the particular family of
wavelets used in the analysis. This is the basic tool needed for
studying time series of $N$ data points via wavelets. Moreover, we
stress the fact that MODWT is based on a $N\log(N)$ computational
algorithm, a fact that makes MODWT as fast as the Fast Fourier
Transform (FFT).

In this paper we make use of the 8-tap Daubechies \textit{least
asymmetric} scaling and wavelet filter (LA8) that looks like the
Mexican Hat wavelet (the second derivative of a Gaussian
function), but which is also weakly asymmetric. The LA8 filters
are characterized by smoothness, relative compactness, and
approximate zero phase properties. In the excellent book by
Percival and Walden \cite{percival} the reader can find all the
mathematical details. For the purpose of this paper, it is
important to have in mind one of the important properties of the
MODWT: the Wavelet Multiresolution Analysis (WMA). It is possible
to prove that given an integer $J$ such that $2^{J}<N$, where $N$
is the number of data points, the original time series represented
by the vector ${\bf X}$ can be decomposed into a smooth part plus
details as follows:
\begin{equation}  \label{decomw}
{\bf X}=S_{J} + \sum _{j=1}^{J} D_j~,
\end{equation}
with the quantity $S_{j}$ generated by the recursion relation
\begin{equation}  \label{decomrel}
S_{j-1}= S_{j} + D_{j}~.
\end{equation}
The detail $D_j$ of Eq. (\ref{decomw}) represents changes on a scale that
 include fluctuations with a period between $2^j$ and $2^{j+1}$ time units,
 while the curve $S_{J}$ smooths fluctuations with a period shorter
 than $2^{j+1}$ time units. We term {\em residuals}
the rough parts of the signal, that is, the quantities with the
smooth average removed
\begin{equation}  \label{decomwR}
R_{J}={\bf X}-S_{J} = \sum _{j=1}^{J} D_j~.
\end{equation}
It is then evident that we can interpret the residual $R_{J}$'s as
fluctuations around the smooth curve $S_{J}$.


\section{Wavelet analysis of the data}

To emphasize the utility of WMA we show, in Figs. 3 and 4, the
properties of teen birth data at various resolutions. These
figures stress the great utility of analyzing times series of
complex systems through the MODWT. The wavelet sensibility to the
local changes of the signal allows us to easily extract an amount
of information impossible to obtain by using, for example, the FFT
that averages the changes of the entire signal at each Fourier
Transform frequency. Fig. 3 shows the smooth curves  and $S_5$,
$S_6$, $S_7$  and  $S_8$ of the births to teenagers in the years
1964-1999; these curves smooth all fluctuations with period
shorter than 64, 128, 256, 512 days, respectively. Fig. 4 shows
the detail curves $D_2$, $D_4$, $D_7$ and $D_8$  of the births to
teenagers in the years 1964-1999; these detail curves capture the
fluctuations of the signal within the scale range of [4:8],
[16:32], [128:256] and [256:512] days, respectively.

\subsection{Comments on the results of the WMA applied to the data}

The smooth curve $S_8$  is enough to emphasize the changes that
occur from year to year. The birth rate is shown in Fig. 3 to be
almost constant during the years from 1964 to1968 with an average
of 113 births per day; during the years 1968-1971 the average
birth rate increases from 113 to 130; the 1971-1978 time period is
characterized by a slow decrease of birth rate from 130 to 125; in
the 1978-1983 time range we see an increase from 125 to 140 births
per day; in the years from 1983 to 1987 there is a decrease from
140 to 127 births per day; finally, the years from 1987 to 1999
are characterized by a steady increase of the birth rate from 127
to 155. The smooth curve $S_7$ shows the annual biases of the
data. There are evident annual cycles characterized by different
amplitude. The details $S_6$ and $S_5$  show complex shapes that
will be studied in detail in Sec. 4B.

The detail $D_8$ in Fig. 4 shows the annual fluctuations. They
have amplitudes (number of births) between 5 and 10; the years
1969, 1973, 1990 and 1999 are characterized by the small amplitude
of 5, whereas the years 1971, 1985 and 1994 are characterized by
the large amplitude of 10. The details $D_4$ and $D_7$  in Fig. 4
are characterized by a mean amplitude of 4 and the details $D_5$ and $D_6$
  are even weaker, with average amplitude of 2. The
details $D_1$, $D_2$ and $D_3$  are characterized by large
amplitude in the range from 5 to 35, a fact that proves that the
largest change of the birth rate occurs during the week. It is of
special interest to notice that the fluctuation intensities of the
detail curve $D_2$ and, in part, of $D_3$, that are almost
constant in the earlier years, show distinct signs of a steady
increase during the years 1990-1999. This property is already
evident in the data illustrated in Fig. 1. The results of Fig. 4
establish that this effect has a weekly origin because the detail
curve $D_2$ represents changes on a scale compatible with the
weekly cycle, that is, between 4 and 8 days. This might be related
to either changes of hospital staff or to the increasing tendency
of health care providers to induce a delivery on particular days,
primarily during the week days, Monday through Friday.

\subsection{Annual patterns in birth and conception rates}

In this subsection we illustrate how to obtain birth and
conception rates during the year by means of the wavelet smoothing
filtering. The main problem to solve is that the annual cycle
fluctuates around a local bias changing from year to year. The key
idea we apply is to determine and detrend this local bias from the
original data and, then, evaluate the resulting annual
fluctuations. Of course, there are many ways to define this local
bias. To justify our choice, we have recourse to Fig. 5a, where we
observe the smooth curves $S_4$  and $S_9$ in the 1982-1999 time
period. These smooth curves smooth fluctuations with a period
shorter than 32 and 1024 days, respectively. We note that the
smooth curve $S_9$ is the curve of lowest-order averaging a period
larger than one year.  Figs. 5b and 5c illustrate the effect of
detrending $S_9$ from the original data and from $S_4$,
respectively. It is evident that this detrending yields
fluctuations around an almost constant and vanishing mean value.

To obtain the birth rate during the year, we use the data from Fig. 5b and
for each day of the year we evaluate the corresponding mean value over the
36 years of the 1964-1999 time period. The results are shown in Fig. 6. The
months with the lowest birth rate are April and May, and those with the
highest are August and September. There is a fast growth of the birth rate
in the months of June and July, whereas the birth rate is almost constant in
October, November and December. Fig. 6 also shows sharp spikes in
correspondence to important holidays: the 1st and 2nd of January, the 4th of
July, the first week of September and, finally, the 25th and 26th of
December. This is likely due to the fact that health care providers schedule
deliveries to avoid holidays when hospitals have limited staff.

A rigorous calculation of the conception rate throughout the year
is difficult to make, but it is possible to estimate it. A good
estimation can be done by assuming that the conception of a baby
takes place about 38 weeks before the delivery. This period is
determined by considering that the traditional duration of
pregnancy. The {\it gestational age}, dates from the first day of
the last menstrual period, an average of 40 weeks prior to
delivery. The pre-ovulatory or follicular phase of the ovarian
cycle, that ends with the ovulation of an oocyte ready for the
conception, is 2 weeks, see Refs. \cite{baby1,baby2}. However, 38
weeks is the mean value, and periods ranging from 36 to 40 weeks
from conception to delivery are still considered to be normal. For
this reason, the details about the variability of the births upon
a temporal scale shorter than 14 days should not be correlated
with the conception rate. We can take this fact into account by
considering the smooth curve $S_4$ (Figs. 5a) that averages the
data upon a wavelet period of 16 days as a reasonable estimate of
error related to the conceptions. Then we apply a procedure
analogous to that used to derive the results of Fig. 6. We use the
data of Fig. 5c, namely the smooth curve $S_4$, with the smooth
curve $S_9$ detrended from it, and for each day of the year we
average over the 36 years of the 1964-1999 time period. Finally,
we shift back the time of 38*7=266 days and, in Fig. 7, plot the
result with the corresponding error bars.

Fig. 7 shows that the lowest rate of conception occurs during the months of
July and August. There is a conception rate increase during the fall that
reaches a first peak in the second half of November. This may be related to
the Thanksgiving holidays. There is a second sharper peak in the second half
of December, corresponding to the Christmas holidays and the end of the fall
school semester. In January the conception rate decreases and from February
to May is almost constant. There is, however, an increase of the conception
rate in the middle of March that may be related to the spring break. Fig. 7
shows clearly that the conception phenomenon may be related to annual
temperature as well as to holidays and the school calendar. Moreover, in
Ref. \cite{newpattitsalis} the interested reader will find a figure
equivalent to Fig. 7 for the conception rate of married and unmarried teens.
The latter figure clearly shows that unmarried teens are influenced by the
one half year school cycle much more than are married teens.

\section{Scaling analysis}

Scale invariance has been found to hold empirically for a number of complex
systems \cite{2Mandelbrot} and the correct evaluation of the scaling
exponents is of fundamental importance to assess if universality classes
exist \cite{stanley}. A widely used method of analysis of complexity rests
on the assessment of the scaling exponent of the diffusion process generated
by a time series (see, for instance, Refs. \cite
{solarflares,thermodynamicofsocialprocesses,scalingdetection,dfa,dea3}).
According to the prescription of Ref. \cite{dfa}, we interpret the numbers
of a time series as generating diffusion fluctuations and we shift our
attention from the time series to the probability density function (pdf) $%
p(x,t)$, where $x$ denotes the variable collecting the fluctuations and $t$
is the diffusion time. In this case, if the time series is stationary, the
scaling property of the pdf of the diffusion process takes the form
\begin{equation}  \label{scafun12}
p(x,t) = \frac{1}{t^{\delta}}~F\left( \frac{x}{t^{\delta}}\right),
\end{equation}
where $\delta$ is a scaling exponent.

The authors of Ref. \cite{scalingdetection} establish that DEA, a
method of statistical analysis based on the Shannon entropy of the
diffusion process \cite{thermodynamicofsocialprocesses} determines
the correct scaling exponent $\delta $ even when the statistical
properties, as well as the dynamic properties, are anomalous. The
other methods usually adopted to detect scaling, for example the
detrended fluctuation analysis \cite{dfa}, are based on the
numerical evaluation of variance. Consequently, these methods
detect a power index, called $H$ by Mandelbrot \cite{2Mandelbrot}
in honor of Hurst, which might depart from the scaling $\delta $
of Eq. (\ref {scafun12}). These variance methods produce correct
results in the Gaussian case, where $H=\delta $
\cite{scalingdetection}, but fail to detect the correct scaling of
the pdf, for example, in the case of L\'{e}vy flight, where the
variance diverges, or in the case of L\'{e}vy walk, where $\delta
$ and $H$ do not coincide, being related by $\delta =1/(3-2H)$
\cite {scalingdetection}. We refer the interested reader to Ref.
\cite {scalingdetection} for further mathematical details.

The Shannon entropy for the diffusion process at time $t$ is defined by
\begin{equation}
S(t) = - \int p(x,t) ~\ln [p(x,t)] ~dx.  \label{entropy}
\end{equation}
If the scaling condition of Eq. (\ref{scafun12}) holds true, it is easy to
prove that
\begin{equation}  \label{scafun14}
S(t) = A + \delta~ \ln(t) ,
\end{equation}
where,
\begin{equation}  \label{ainthecontinuouscase}
A \equiv -\int_{-\infty}^{\infty} dy \, F(y) \, \ln [F(y)],
\end{equation}
and $y = x/t^{\delta}$. Numerically, the scaling coefficient $\delta$ can be
evaluated by using fitting curves with the form $f_S(t)=K + \delta \ln (t)$
that on a linear-log scale is a straight line.

\subsection{The diffusion algorithms}

Let us consider a sequence of $N$ numbers
\begin{equation}
\xi_{i} , \quad i = 1, \ldots , N.  \label{thesequenceundestudy}
\end{equation}
The goal is to establish the possible existence of scaling, either normal or
anomalous within this sequence. First, let us select an integer number $t$,
fitting the condition $1 \leq t \leq N .$ We adopt the symbol $t$ for this
integer number because it plays the role of diffusion time. For any given
integer time $t$ we can find $N - t +1$ sub-sequences defined by
\begin{equation}
\xi_{i}^{(z)} \equiv \xi_{i + z}, \quad with \quad z = 0, \ldots , N-t.
\label{multiplicationofsequence}
\end{equation}
For any of these sub-sequences we build up a diffusion trajectory, $z$,
defined by the position
\begin{equation}
x^{(z)}(t) = \sum_{i = 1}^{t} \xi_{i}^{(z)} = \sum_{i = 1}^{t} \xi_{i+z}.
\label{positions}
\end{equation}

The DEA is applied to these trajectories according to the following
algorithm: (1) we partition the $x$-axis into cells of size $\epsilon$ and
label the cells with an integer index $j$; (2) we count how many
trajectories are found in the same cell at a given time $t$ and denote this
number by $N_{j}(t)$; (3) we use this number to determine the probability
that a particle can be found in the $j$-th cell at time $t$, $p_{j}(t)$, by
means of
\begin{equation}
p_{j}(t) \equiv \frac{N_{j}(t)}{(N-t+1)} .  \label{probability}
\end{equation}
At this stage the entropy of the diffusion process at the time $t$ becomes:
\begin{equation}
S(t) = - \sum_{j} p_{j}(t) ~\ln [p_{j}(t)].  \label{entropy}
\end{equation}
The easiest way to proceed with the choice of the cell size, $\epsilon$, is
to assume it to be a fraction of the square root of the variance of the
fluctuation $\xi_i$, and consequently to be independent of $t$. If the
scaling condition, Eq. (\ref{scafun12}), holds true, Eq. (\ref{scafun14}) is
fulfilled by Eq. (\ref{entropy}) and the scaling exponent $\delta$ may be
determined with a fitting procedure.

\subsection{Data analysis}

The numerical results obtained by applying the DEA to three data
sets are shown in Fig. 8 : (i) the original teen birth time series
(black curve); (ii) the time series derived from the original by
detrending the annual cycle from it (gray circle curve); (iii) the
time series derived from the original
by detrending from it not only the annual cycle but also the linear ramp $%
y=a+b(t-1964)$ discussed in Sec. 2 (gray star curve). The annual
cycle is identified with the details $D_7$ and $D_8$, and
consequently the gray circle curve refers to the original data
with the details $D_7$ and  $D_8$ detrended from them. We would
like to stress an important property of wavelet detail curves.
Contrary to the sine and cosine functions used in the Fourier
transform, the wavelet details are characterized by a wide
spectrum. Because of this property, removing a wavelet detail from
the original data reduces the spectral intensity of the
frequencies in a region of the spectrum without eliminating these
frequencies completely. This means that removing the details $D_7$
and $D_8$ may be considered a {\it light} filtering procedure that
smoothes the annual cycle but leaves some background noise in that
spectral region. This will not completely disrupt the correlations
of the noise in that spectral region.

 The diffusion entropy, $S(t)$, of the
original data (black curve in Fig. 8) shows an interesting shape.
In fact, it increases with a slope faster than that of the random
noise ($\delta =0.5$) but does not present a real scaling behavior
but a periodic one. The diffusion entropy of the original data
increases and decreases giving origin to cusps with a periodicity
of one year. These cusps are due to the periodic annual
convergence of the distinct trajectories (\ref {positions}) at the
completion of the annual cycle that causes a reduction of the
entropy. In the next sections we will discuss in detail this
behavior.

Two straight lines corresponding to $S(t)=1.35+0.80\ln (t)$ and to $%
S(t)=0.93+0.81\ln (t)$ to emphasize the accuracy of the scaling
detection for the two detrended  datasets are also shown in Fig.
8. We see that the
first straight line fits the gray circle curve with a scaling coefficient $%
\delta =0.80$ very well for a long temporal range that lasts up to
few years. The gray circle curve refers to the residual noise
obtained from the original data with the details $D_7$ and $D_8$
detrended from them. We also note the important fact that this
straight line intercepts the cusps of the black curve that
corresponds to the DEA of the original data. As is shown in the
next section, the goodness of the fit of the gray circle curve
means that detrending the annual periodicity from the original
data allows us to discover that the background noise of the data
fulfills the scaling condition expressed by Eq. (\ref{scafun14}).

The gray star curve of Fig. 8 is the DEA of the residual noise obtained from
the original data with the details $D_7$ and $D_8$ and the linear fitting ramp $%
y=a+b(t-1964)$ discussed in Sec. 2 detrended from them. This is only a test
for the stability of the scaling $\delta =0.80$ measured by fitting the gray
circle curve. In fact, it may happen that linear ramps disrupt the fractal
properties of the signal, cause a superdiffusion of the trajectories (\ref
{positions}) and induce a pseudo-scaling. However, this takes place only if
the biases induced by those ramps are large compared to the natural fractal
persistencies of the signal. Fig. 8 shows that the gray star curve is very
well fitted for a long period of time (from 100 to 1500 days) by a straight
line with $\delta =0.81\pm 0.02$ that is practically parallel to the first
straight line with $\delta =0.80\pm 0.02$. This establishes that the scaling
of the gray circle curve, $\delta =0.80$, is not an artifact of the linear
ramp given by the fitting average $y=a+b(t-1964)$ discussed in Sec. 2. The
long bias showed by the data should be considered compatible to a long
fractal persistency of the signal. Finally, Fig. 8 shows that the transition
period ($t<100$) to the scaling condition that the gray star curve requires,
seems to be due to the emergency in the entropic analysis of the strong
weekly cycle that disrupts the scaling at short time. In Sec. 7 we will
discuss in more detail why short cycles of the signal may emerge in the
entropic analysis after a detrending procedure.

The scaling of the gray circle curve of Fig. 8 is an interesting result. In
fact, it is an important step towards the solution of a problem raised by
earlier work. In the earlier investigation of Ref. \cite
{thermodynamicofsocialprocesses} the question was raised as to a
significantly large scaling being an artifact of the adoption of a technique
without detrending. The doubt was also raised that, after applying a
detrending procedure, the resulting scaling might be very close to that of
ordinary Brownian motion. This would significantly weaken earlier claims\cite
{morememory} that the teen birth phenomenon is a complex process yielding
anomalous scaling. We see from Fig. 8 that the scaling emerging from the
detrending procedures corresponding to the fitting straight line of Fig. 8,
is $\delta=0.80\pm0.02$, and it is slightly weaker than that produced by a
tangent to the black curve, the slope of this tangent being very close to
the prediction of techniques with no detrending. However, we see that the
detrended scaling is still significantly larger than the scaling of ordinary
random noise ($\delta = 0.5$). Furthermore, this scaling ranges up to $t =
2000$ days, namely a wide time range including five annual
periodicity-induced entropy regressions to local minima. This 
result fully supports the perspective proposed by the authors of Ref. \cite
{morememory}, that the teen birth phenomenon is a complex process. In fact,
the careful detrending of annual periodicity makes an anomalous scaling
emerge and last for a very extended time. The straight line of Fig. 2
corresponds to the scaling condition with $\beta=2H-1=0.60$, assuming $%
H=\delta=0.80$. The scaling of the straight line appears consistent with the
scaling behavior of the spectrum of the background noise of the data for a
large period from some months to few years.

\section{A physical model for periodicity-induced breakdown of the scaling
condition}

We have seen that the yearly periodicities distort the scaling condition
and, therefore, must be removed from the data. In this section we study this
distortion assuming that the hidden scaling regime is due to the correlation
of fractional Gaussian noise. We find that the model of this section yields
effects that are very similar to those illustrated in Fig. 8. In conclusion,
we assume that a random sequence by itself would generate a diffusion
process with scaling, namely, fulfilling the scaling prescription of Eq. (%
\ref{scafun12}). We shall see that the model for non-stationarity adopted in
this section affords a satisfactory account for the qualitative properties
of the numerical results, and especially for the very slow attainment of the
scaling regime. We find that the scaling regime is reached throughout by a
slow relaxation process characterized by inverse power law behavior.


We assume that the random component of the signal under study, $\{\xi_i\}$,
generates a diffusion process whose pdf, $p_s(x,t)$, obeys the scaling
prescription of Eq. (\ref{scafun12}), that is:
\begin{equation}  \label{scafun122}
p_s(x,t) = \frac{1}{t^{\delta}}~F\left( \frac{x}{t^{\delta}}\right),
\end{equation}
where $t$ is the diffusion time. In the numerical calculation we shall
assume $\{\xi_i\}$ to be ordinary Brownian noise, thereby setting $\delta =
0.5$ and assigning a Gaussian form to the function $F(y)$. The theoretical
analysis, however, is more general, and, as we shall see, seems to leave
room for cases where the lifetime of the periodicity-induced non-stationary
condition is infinite. Then we add to these data a sinusoidal function with
period $T$ and amplitude $A$. In conclusion, the sequence that we propose as
a model for periodicity-broken stationarity is given by
\begin{equation}  \label{newseq}
\zeta_i=\xi_i + A\sin\left(\frac{2\pi ~i}{T} \right)~.
\end{equation}

Let us apply to the sequence (\ref{newseq}) the diffusion algorithm of Sec.
5A. The first step consists of creating many distinct trajectories, each
being denoted by the subscript $z$. These trajectories at time $t $ yield
the position corresponding to the prescription of Eq. (\ref{positions}) and
reads in this case as follows:
\begin{equation}
x^{(z)}(t) = \sum_{i = 1}^{t} \xi_{i+z}+ \sum_{i = 1}^{t}A\sin\left[\frac{%
2\pi ~(i+z)}{T} \right]~.  \label{positions2}
\end{equation}
Of course, the first and second terms on the right-hand side of
Eq. (\ref{positions2}) correspond to the contributions generated
by the random and sinusoidal component, respectively. In the limit
case of very large time periods $T$, it is possible to adopt for
the second term the continuous-time approximation. This is
realized by replacing the second sum of Eq. (\ref {positions2})
with an integral. The integration over the index $i$, thought
of as a continuous variable, makes the discrete sum equivalent to the value $%
y$ given by:
\begin{equation}  \label{valinteg}
y \equiv g(z,t) \equiv \frac{AT}{\pi}\sin\left(\frac{\pi ~t}{T} \right)
\sin\left(\frac{2\pi ~z}{T}+\frac{\pi ~t}{T}\right)~.
\end{equation}
This component oscillates from positive to negative values, whose maximum
intensities, $y_{max}$ for the positive values and $-y_{min}$ for the
negative values, are given by:
\begin{equation}  \label{yminmax}
y_{max}=-y_{min}=\frac{AT}{\pi } ~\|\sin \left( \frac{\pi ~t}{T}\right)\|~.
\end{equation}

Using the fact that the random component is totally independent of the
deterministic component, and viceversa, we determine that the resulting pdf,
$p(x,t)$, is given by the convolution
\begin{equation}  \label{finpdf}
p(x,t)=\int\limits_{y_{min}}^{y_{max}} ~p_{s}(x-y,t)~p_{d}(y,t)~dy~.
\end{equation}
The choice we make for the random component is, as mentioned earlier, that
of ordinary Brownian motion, thereby yielding
\begin{equation}  \label{gaussdistr}
p_{s}(x,t)=\frac{1}{\sqrt{2\pi t}} ~\exp \left(- \frac{x^2}{2t} \right)~.
\end{equation}
The definition of the deterministic component $p_{d}(y,t)$ requires some
additional work. From Eq. (\ref{valinteg}) we see that the probability that
the diffusing variable realizes a value in the interval $[y, y + dy]$ is
given by the geometric probability
\begin{equation}  \label{prodely}
p_d(y,t) dy = \frac{dy \sqrt{1+ (dz /dy)^{2}}}{\int\limits_{0}^{T/2}dz \sqrt{%
1+ (dy/dz)^{2}}},
\end{equation}
the integration over $z$ being confined to one half time period. The meaning
of Eq. (\ref{prodely}) is that the probability of finding the variable in a
given interval of size $dy$ is the ratio of the size of the length of the
portion of the curve $y(z)$ corresponding to that interval to the total
length of the same curve corresponding to one half period. The second
derivative of $y$ with respect to $z$ and its inverse are evaluated using
Eq. (\ref{valinteg}). By plugging the two resulting expression into Eq. (\ref
{prodely}), we obtain
\begin{equation}  \label{pdfd}
p_{d}(y,t)=\frac{\sqrt{1+\left[1/f(y,t) \right]^2}}{\int\limits_{0}^{T/2}%
\sqrt{1+\left[g^{\prime}(z,t) \right]^2}~dz}~,
\end{equation}
where
\begin{equation}  \label{funcf}
f(y,t) \equiv 2A\sin \left(\frac{\pi ~t}{T}\right)~\sqrt{1-\left[\frac{\pi ~y%
}{TA\sin \left(\frac{\pi ~t}{T} \right)} \right]^2}~,
\end{equation}
and
\begin{equation}  \label{functg}
g^{\prime}(z,t) \equiv \frac{\partial g(z,t)}{\partial z}=2A\sin \left(\frac{%
\pi ~t}{T} \right)\cos \left(\frac{2\pi ~z}{T} + \frac{\pi ~t}{T}\right)~.
\end{equation}
Note that $f(y,t)$ is nothing but $g^{\prime}(z,t)$ expressed as a function
of $y$, Eq. (\ref{valinteg}) rather than of $z$.

We are now ready for a visual representation of the time evolution of $%
p(x,t) $ stemming from Eq. (\ref{finpdf}) with the earlier described
prescriptions to determine $p_{s}(x,t)$ and $p_{d}(x,t)$. We do the
calculation with the values $T=100$ and $A=1$, and show the results in Fig.
9. We see that in this condition the resulting distribution shows signs of
the presence of the deterministic component $p_{d}(x,t)$. The time evolution
of this component is as follows. At $t=0$ the pdf is a delta of Dirac peaked
at $x=0$. Upon increase of time this peaked initial condition broadens and
its amplitude decreases. The maximum spreading and minimum intensity is
reached at half cycle, then the inverse process begins to recover, till at
the end of the cycle, the initial distribution is again peaked at $x=0$.
Then the distribution width and intensity begin broadening and decreasing
again, and so on, infinitely many times. The presence of the random
component has the effect of making incomplete the regressions to the initial
condition, so that only a portion of the initial intensity is recovered.
Upon increase of time the system tends to the distribution that it would
have if only the random component were present, and the scaling of this
component is recovered. We interpret scaling as a form of thermodynamic
equilibrium \cite{solarflares}, and we see that the presence of the
deterministic component has the effect of slowing the transition from the
initial Dirac delta function to the final scaling regime. This effect can
also be interpreted as an enhancement of the time duration of the
non-stationary condition, which extends to a very large portion of the
observation time. This is made very clear by Fig. 10, where we compare the
intensity of $p(x,t)$ to the intensity of $p_{s}(x,t)$, namely the values
that these two distributions have at $x=0$. We see that these intensities
coincide only at the regression times of $p_{d}(x,t)$, and that it takes a
very extended time for $p(x,t)$ to become identical to $p_{s}(x,t)$.

In conclusion, the existence of a cyclical deterministic component has the
effect of creating a non-stationary condition with a very slow regression to
the scaling regime. It is possible to express analytically the form of this
relaxation process, in a condition more general than that examined by the
numerical calculations. Let us define the quantity
\begin{equation}
k(t)=p_{s}(\tilde{x},t)-p(\tilde{x},t)~,  \label{kkl}
\end{equation}
which measures the difference between the scaling distribution of Eq. (\ref
{scafun122}) and the convoluted distribution (\ref{finpdf}) at the position $%
x=\tilde{x}$ where the distribution (\ref{scafun122}) has its maximum. This
means that this analysis applies also to cases different from the symmetric
Gaussian case, to which we are referring our numerical check. In the
Gaussian case here under numerical study, we have $\tilde{x}=0$. In general
we set the condition
\begin{equation}
\left. \frac{\partial }{\partial x}p_{s}(x,t)\right] _{x=\tilde{x}}=0~.
\label{maxcom}
\end{equation}
We refer $p(x,t)$ of Eq. (\ref{finpdf}) to $x=\tilde{x}$; use for $p_{s}(%
\tilde{x}-y)$ the scaling condition of Eq.(\ref{scafun122}) and expand it
into a Taylor series with respect to $y=0$. Since the size of the
deterministic distribution $p_{d}(y,t)$ remains below a given maximum value,
we can safely make the assumption that, at least for suitably large values
of $t$, the values of $y/t^{\delta }$ are so small as to make it legitimate
to truncate the Taylor series expansion of $p_{s}(\tilde{x}-y)$ at the order
corresponding to its first non-vanishing derivative. We denote this order
with the symbol $\nu $. Then, we insert the resulting expression for $p(x,t)$
into Eq. (\ref{kkl}). The contribution corresponding to the zero-th order
term of the Taylor expansion cancels with the second term on the right hand
side of Eq.(\ref{kkl}), and from the term corresponding to the first
non-vanishing derivative of $p_{s}(\tilde{x}-y)$, we easily derive the
following inverse power law expression
\begin{equation}
\lim_{t\to \infty }k(t)\propto \frac{1}{t^{\delta +\nu \delta }}.
\label{kkklas}
\end{equation}
Let us focus on cases where the first non-vanishing derivative is that of
second order: $\nu=2$. This is the condition fulfilled by the Gaussian case of our
numerical treatment. In this case, we immediately find from Eq.(\ref{kkklas}%
) that the inverse power index of this slow decay is $3\delta $. In the numerical
example, we are studying $\delta =0.5$. Consequently we expect in that case
that the inverse power law relaxation has the index $1.5$. More generally,
in the subdiffusional case $\delta <1/3$ the relaxation of the
non-stationary condition is not integrable, and the lifetime of the
non-stationary condition is infinite. The study of this interesting case is
left for future investigation. Here we limit ourselves to observing that in
the Gaussian case, here under study, $\tilde{x}=0$. At the regression times $%
t=T/2+nT$, with $n=1,2,3,...$, the distribution $p_{d}(x,t)$ has its maximum
spreading and at $x=0$ the pdf $p(x,t)$ reaches its local minima as shown in
Fig. 10. Therefore, we expect in this case very good agreement between
theory and numerical calculations. This is fully confirmed by Fig. 11, where
the theoretical inverse power law relaxation with index $1.5$ fits the
numerical results over many decades.

What about the results of DEA applied to this non-stationary condition? This
is illustrated by Fig. 12, showing the DEA applied to the sequence $%
\{\zeta_i\}$ of Eq. (\ref{newseq}) (solid curve) and to the
Gaussian component $\{\xi_i\}$ alone (dashed curve). Of course we
are considering the sequence $\{\zeta_i\}$ as corresponding to the
experimental data of this paper, and the sequence $\{\xi_i\}$ as
obtained from this experimental sequence by detrending from it the
seasonal periodicity. We see that the qualitative behavior of Fig.
8 is very satisfactorily reproduced. This figure shows that the
initial slope of the full curve (mimicking that emerging from the
experimental data) does not have anything to do with the scaling,
corresponding to the slope of the dashed line. This property
suggests that we interpret the straight line of Fig. 8, concerning
gray circle curve obtained by the data with the wavelet details
$D_7$ and $D_8$  detrended from them, as the scaling of the teen
birth process under study.

\section{Multiresolution Diffusion Entropy Analysis}

In the previous sections we have introduced a method of analysis, based on
the joint use of Wavelet decomposition and diffusion approach to scaling,
resting on DEA. The main idea was to detrend from the data a part of the
signal, in that case the part related to the annual cycle, and study the
stochastic properties of the reduced signal. In this section we generalize
the method and we refer to it as Multiresolution Diffusion Entropy Analysis,
MDEA.

As shown in the Introduction, Eq. (\ref{noiff}), a signal $\{\xi
_{i}\}$ may always be thought of as the superposition of a
function $f(t)$ plus a noisy component $\{\eta _{i}\}$. The choice
of the function $f(t)$ may be arbitrary. However, there are cases
in which the choice of a particular function $f(t)$ has a
reasonable meaning. For example when it is possible to determine a
component of the signal that has an understandable deterministic
cause. In the previous sections, we have analyzed the data
detrended of the wavelet details $D_7$ and $D_8$ only because
these details contain the annual and one-half annual cycles that
have a simple biological and sociological explanation. It is also
common in the scientific literature to identify the function
$f(t)$ through a smoothing process of the data. The smoothing
procedure helps a visual interpretation of the general trend of
the data as we have studied in Sec. 4A about the wavelet smooth
curve $S_8$. However, the smoothes of a signal do not necessarily
have any real deterministic meaning. Moreover, one should be
warned that the residual noise, $\{\eta _{i}\}$, obtained after
detrending from the data one of its smooth components, may be
characterized by stochastic properties different from those of the
original dataset.

For example, detrending from a fractional Gaussian noise its smooth
component obtained by using a moving average of period T will significantly
deform the fractal correlations of the signal on a time scale larger than T.
The reason is that those long correlations are absorbed into the smooth
curve $f(t)$. Therefore, the residual noise $\{\eta_i\}$ will not be
characterized by the same fractal properties of the original dataset.

An artificial realization of a fractional Gaussian noise with
H=0.80 and its smooth curve obtained with a moving average of
period T=300 is shown in Figure 13a. Note the strong persistences
of this type of fractal noise. Fig. 13b shows the DEA applied to
the original fractal noise (upper curve) and to the residual noise
(lower curve) obtained by detrending the original fractal noise of
the smooth component. The figure shows clearly that the detrending
procedure disrupts the scaling properties of the original signal.
In fact,
while the DEA of the fractal noise yields the correct scaling exponent of $%
\delta =0.80$, the DEA of the residual noise shows a transient
pseudo-scaling of $\delta =0.57$ followed by a $\delta =0$, both
of which are only artifacts of the detrending procedure. In
conclusion, the decomposition of the signal given by Eq.
(\ref{noiff}) is fully justified only when the function $f(t)$
expresses some aspect of the process associated with an
understandable deterministic process.  Instead, when the function
$f(t)$ is not associated with an understandable deterministic
process but, for example, is simply a smoothed part of the signal,
detrending it from the data is not only arbitrary but may be a
disrupting and dangerous procedure as well. The danger concerns
the conclusions about the stochastic properties of the residual
noise that may be artifacts of the decomposition.

There is, however, another case in which the decomposition expressed by Eq. (%
\ref{noiff}) have some justification. If the decomposition of the
signal is regulated by some mathematical theorem, we may refer to
that decomposition as a {\it standard model}. In Sec. 3 we have
seen that the MODWT defines a set of scaling smooth curves, $S_j$,
obtained by the recursion relation (\ref {decomrel}). We may use
this set of smooth curves as a model for the function $f(t)$ of
Eq. (\ref{noiff}) that express the properties of the signal at
different scales and we may study the stochastic properties of the
residual noise to determine a possible interpretation. Fig. 14
shows the DEA of the births to teenagers applied to the wavelet
residuals $R_j$ obtained via Eq. (\ref{decomwR}), where, as in the
earlier sections, $j$ indicates the wavelet scale index. We stress
the fact that each residual contains all details at smaller
scales. Therefore, the WMDA allows us to determine the diffusion
spreading at each time scale, as stemming from the corresponding
details, an important piece of information. The residuals $R_j$
are obtained by detrending from the original data the smooth
curves $S_j$, afforded by the wavelet multiresolution analysis,
see Eq. (\ref{decomwR}). The result of this analysis is
illustrated by Fig. 14. The curves of this figure denote the DEA
of the original data, and the residuals from $R_3$ to $R_8$. These
curves are an illustration of WMDA and indicate that this may be a
useful tool to study the complexity of dynamical systems. Let us
analyze them in more detail.

(i) The straight line refers to $f_{S}(t)=0.5~\ln (t)$, which
becomes a straight line by adopting the linear-log representation.
According to Mandelbrot \cite{2Mandelbrot}, curves with slopes
larger than 0.5 indicate persistent diffusion or superdiffusion,
and curves with slopes smaller than 0.5 indicates antipersistent
diffusion or subdiffusion region. The figure shows that only the
residuals $R_7$, $R_8$ and those of higher order, present a
persistent behavior in the initial time region. This persistence
is mainly due to the periodicity that characterized those wavelet
residuals.

(ii) Each curve is characterized by one leading periodicity. The
dynamical reason for this property is easily accounted for by
noticing that a given periodicity of the data causes a periodic
convergence of distinct trajectories. After an initial spreading,
with a consequent increasing of both variance and entropy, there
is incomplete regression to the initial condition. Since each
curve corresponds to a given scale, the observed processes of
regression correspond to the leading periodicity of that temporal
scale. Any residual $R_{j}$ contains all details at smaller
scales, and, as a consequence, the leading periodicity is not
necessarily related to the wavelet temporal scale $\tau _{j}$ by
simple relations. For example, the  $R_7$ and $R_8$ curves of Fig.
14 clearly show the year and one half year periodicity, whereas
the curve $R_6$ shows the 70-80 days periodicity. The curve $R_5$
is characterized by the weekly cycle and also by a slight 45 days
periodicity. The other curves show the strong weekly periodicity.
In conclusion, Fig. 14 shows that WMDA may be an interesting
complement to the spectral density analysis of Figs. 2 because it
shows the main periodicity that characterizes each scale level.

(iii) Any residual converges to a horizontal line. This is due to the
detrending of the smooth part of the data, $S_j$, that reveal the hidden
periodicities. At the same time, these hidden periodicities imply the
deterministic nature of the signal and consequently yield entropy
saturation. In other words, the fluctuations $\xi_i$ of the residuals data $%
R_j$ can generate trajectories with only a limited spreading. The height of
the horizontal lines measures the information or entropy associated with
each time scale. It is interesting to notice that the shorter the time scale
the faster the transition to saturation. This means that the role of
periodicities becomes more and more important as we decrease the time scale.

Table I records some relevant aspects of the information contained in Fig.
14.

\section{Conclusion}

In this paper we have presented a statistical methodology for analyzing a
complex phenomenon in which a deterministic component and a fractal
correlated noise are superimposed. The goal of our technique is to isolate
and separate a part of the signal with an understandable interpretation
identified by some detail from the scaling part of the signal. Detrending
from the data only some detail components instead of the smooth component
does not distort the mutual influence of the low-frequency and
high-frequency components of the data and preserves the fractal properties
of the background noise. The list of the results and conclusion of this
paper is as follows:

(i) {\em Wavelet analysis of the births to teenagers}. Wavelet
multiresolution analysis is a powerful method for analyzing time series data
by decomposing the signal on a {\it scale-by-scale} basis. In this way we
determine, by means of an easy visual procedure, some interesting properties
of the births to teenagers in Texas. We determined the local biases that
allow us to form an impression of the social changes that have been
occurring over the last 36 years. We assessed that the increasing spreading
of the data during the years 1990-1999 is related to changes in the weekly
patterns of health care delivery. Finally, we determined information about
the birth and conception rates throughout the year, a property that may be
important from a sociological point of view.

For readers to appreciate the importance and the limitations of
this discovery, we draw their attention to two problems when
estimating rates of conceptions on a given day, using only dates
of birth from birth certificates. First, a consistent number of of
teen conceptions, almost 35\% in Texas, does not end in live birth
but in induced or in stillbirth miscarriage. Thus, several teen
conceptions are not reflected on any birth
certificate\cite{patti1}. Second, estimates of premature birth
rates range from $4.4\%$ to $21.5\%$ of all births with even
higher rates of prematurity occurring among minorities and very
young women\cite{patti3}. Using the normal length of gestation,
therefore, can lead to inaccuracy in calculating conception dates
in these cases. With these limitations in mind, we suggest that
the findings regarding conceptions are relevant only to
pregnancies ending in live birth and that while errors in dates of
conception have been minimized by allowing for a two week error in
the length of gestation, the errors may be greater in some cases.
Resolving these problems is beyond the scope of this study,
however.

(ii) {\em Periodicity and scaling}. We established that the DEA may be
affected only by strong and long periodicities like the annual cycle. This
scaling is not influenced by the details that contain the shorter
periodicities, such as the weekly cycle. The annual cycle is responsible for
a non-stationary condition that makes it difficult to detect the underlying
scaling. However, if the effect of the annual periodicity is detrended from
the data, a stationary condition with a well-defined scaling index emerges
at least for a long period of time.

(iii) {\em Residuals and periodicities}. We showed that weak periodicities
emerge at the level of residuals, namely the portions of the signals
obtained after detrending the smooth parts. The DEA of residuals saturates,
thereby implying that after a given time there is no further information
increase. We can, therefore, conclude that there should be no concern about
a possible influence of periodicities on subsequent scaling.

(iv) {\em Joint use of entropy and decomposition}. The benefit enjoyed by
the joint use of DEA and wavelets is evident. The wavelet decomposition
generates a set of new time series, corresponding to tuning the wavelet
microscope to a given time scale. By systematically varying these components
it is possible to distinguish a non-stationary component from the stationary
component of the signal. The DEA establishes the scaling of the stationary
component, if it exists, and makes it evident why periodicities set an upper
limit on entropy increase.

The scaling of the stationary part suggests that, after the detrending of
the deterministic annual cycle, the births to teenagers are correlated
noise, that is, teen birth phenomenon is characterized by the existence of
some type of cooperative effects. These may be related to demographic and
political changes and to the anomalous changing of human fashions whose
persistence last from some months to a few years.

--------\newline
{\ {\large {\bf Acknowledgment:}}}\newline
The authors  gratefully acknowledges Dr. P. Grigolini and Dr. P. Hamilton for useful conversations.  N.S. thanks the Army
Research Office for support under grant DAAG5598D0002.



\onecolumn

\newpage

\begin{figure}[tbp]
\epsfig{file=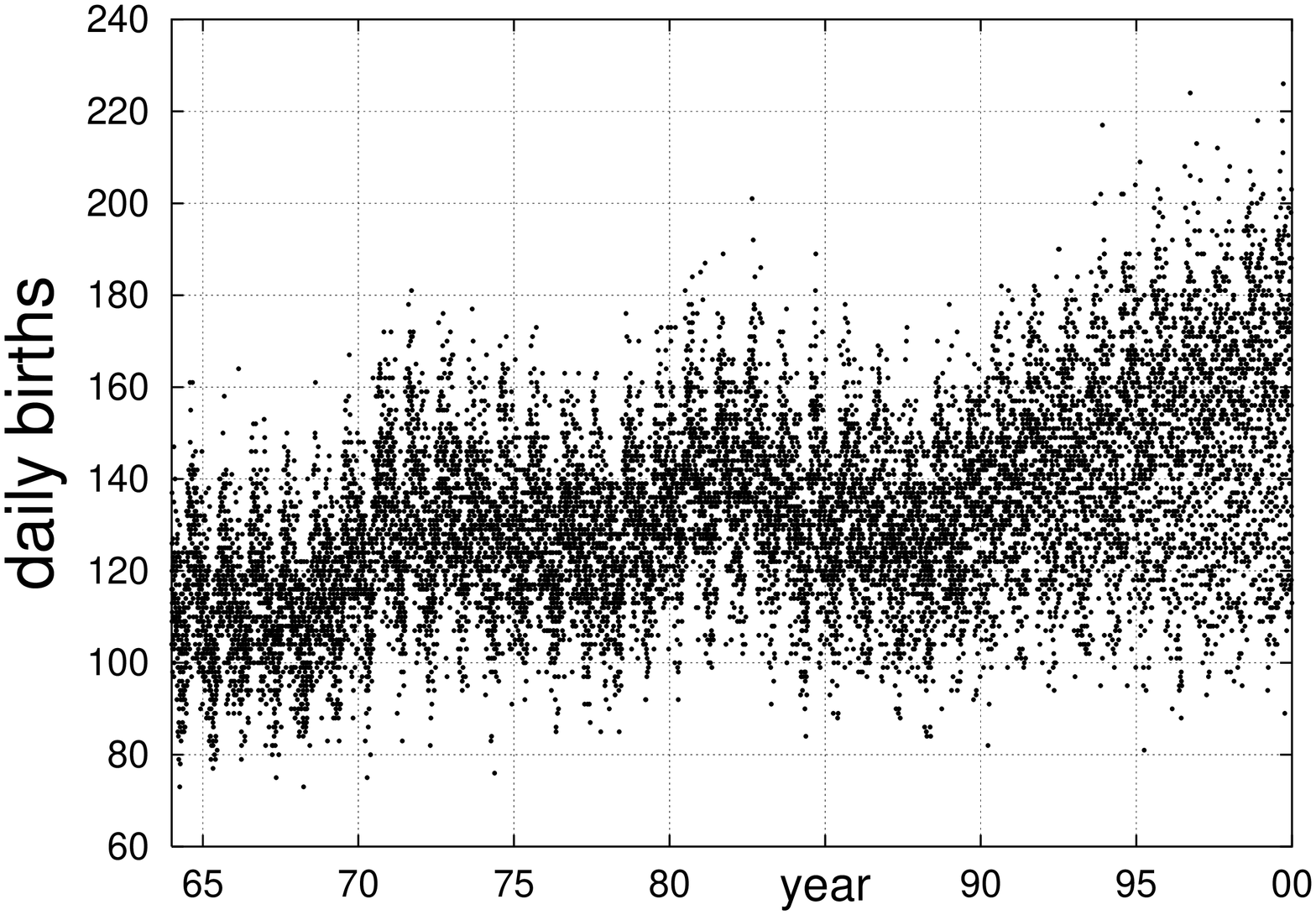,height=15cm,width=10cm,angle=-90} \caption{
Births to teenagers during the years 1964-1999 in Texas.}
\end{figure}

\newpage

\begin{figure}[tbp]
\epsfig{file=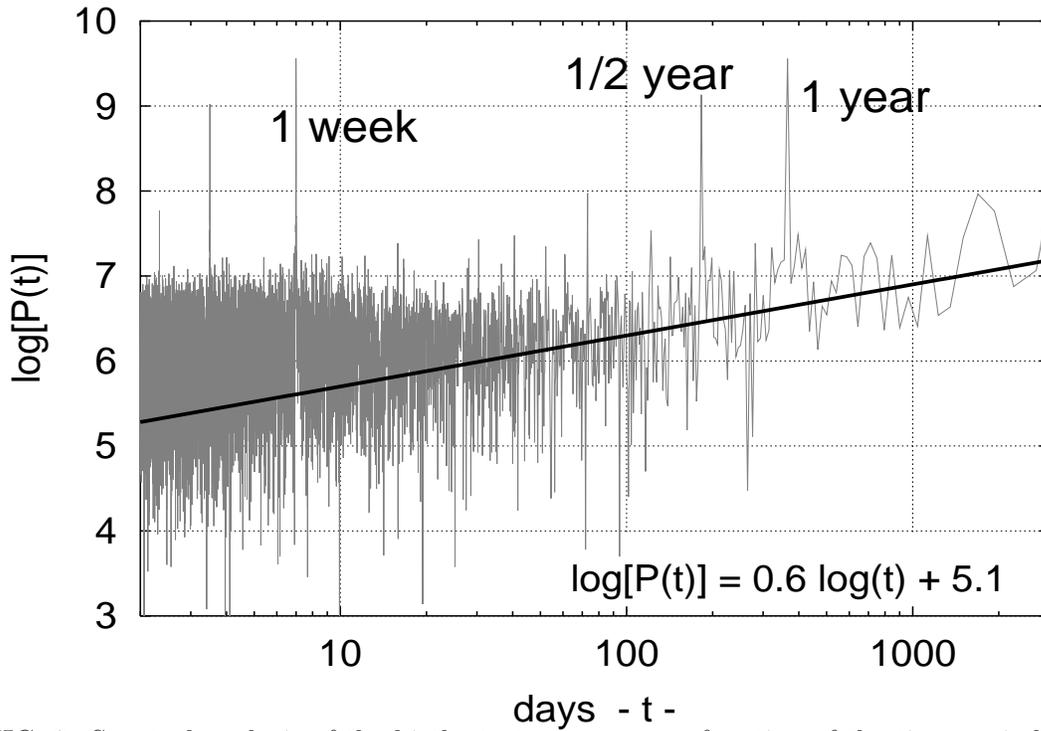,height=15cm,width=10cm,angle=-90}
\caption{Spectral analysis of the births to teenagers as a
function of the time period, measured in days. From the right to
the left the periods revealed by this analysis are: 3.5 days, 7
days, 73 days, 183 days, 365 days. The background noise exhibits a
scaling behavior similar to that of a fractional Gaussian noise at
least for a large period of time, from one or two months to 8
years. The straight line corresponds to the scaling condition with
$\beta=2H-1=0.60$ assuming $H=\delta=0.80$ measured in Sec. 5. }
\end{figure}

\newpage
\begin{figure}[tbp]
\epsfig{file=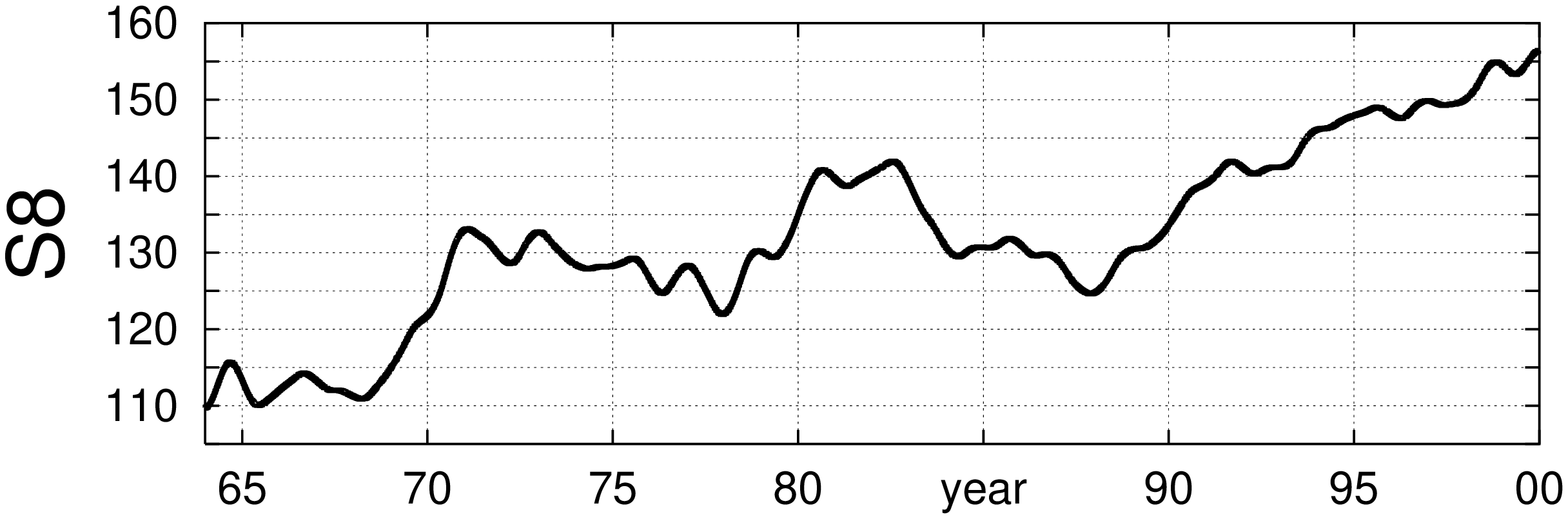,height=15cm,width=4.5cm,angle=-90}

\epsfig{file=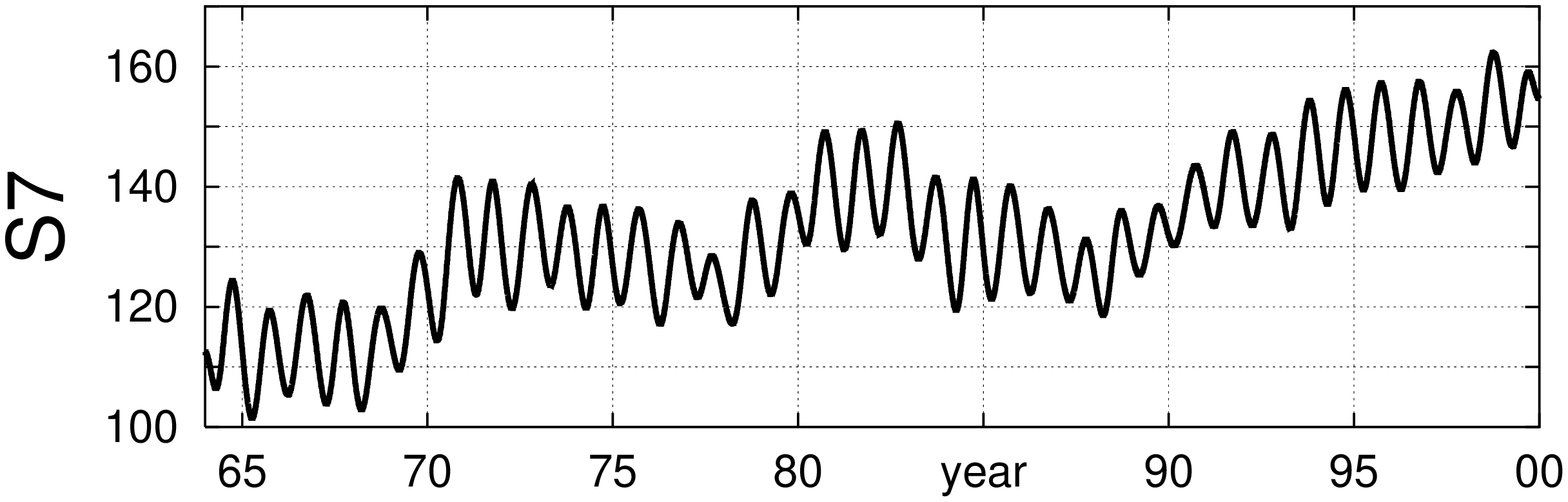,height=15cm,width=4.5cm,angle=-90}

\epsfig{file=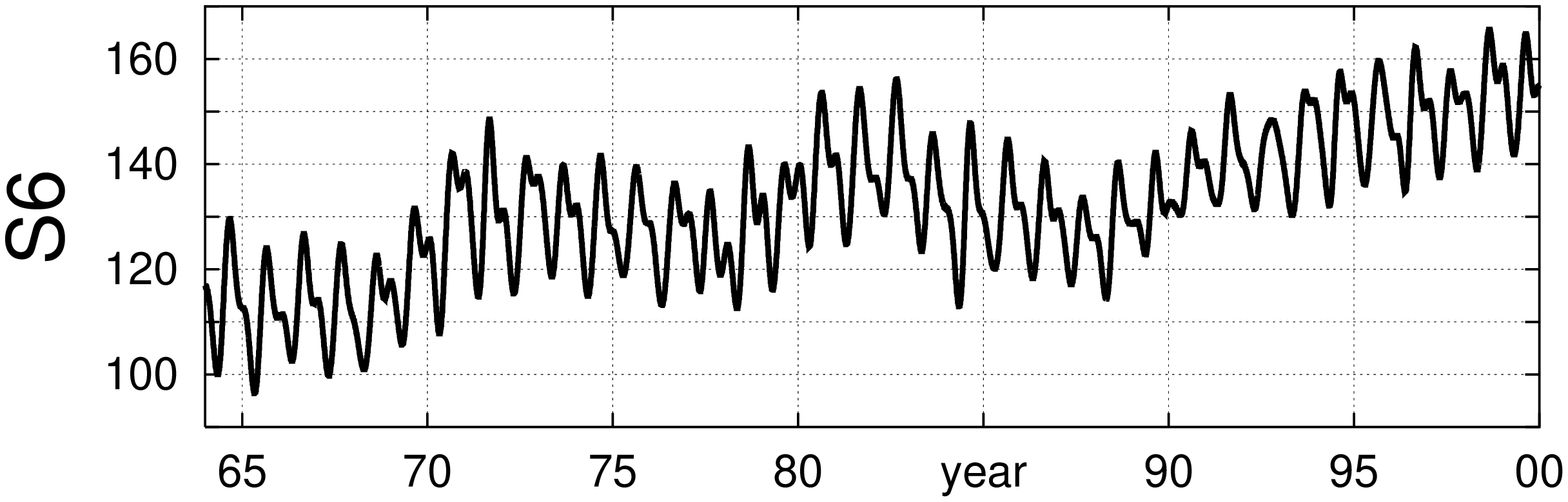,height=15cm,width=4.5cm,angle=-90}

\epsfig{file=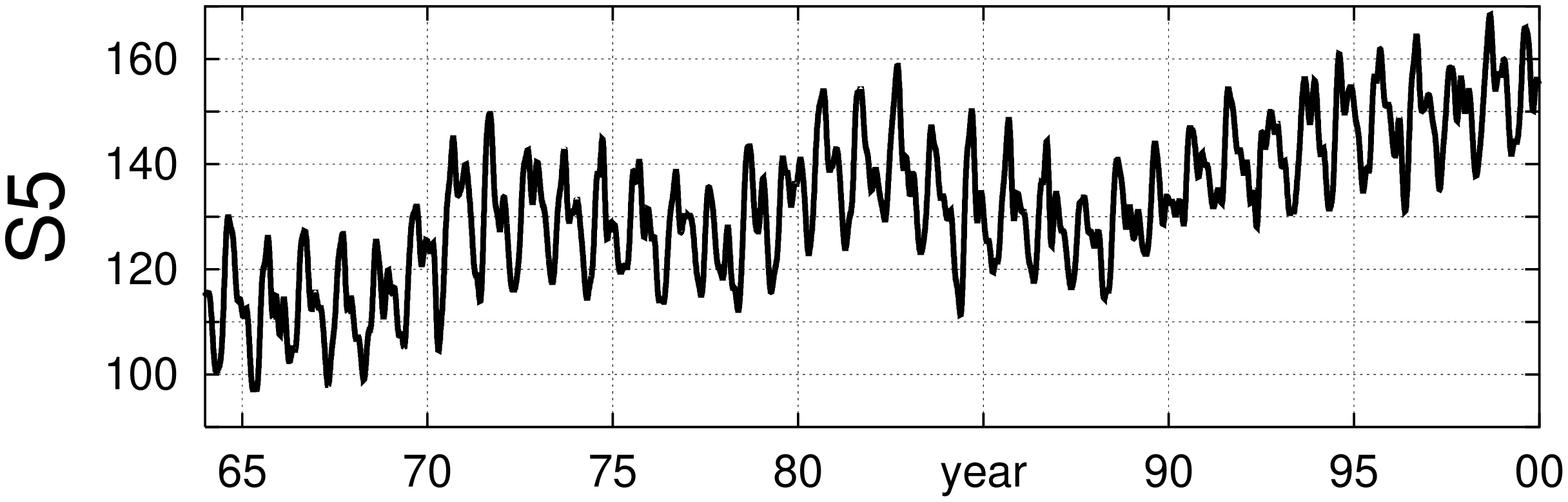,height=15cm,width=4.5cm,angle=-90}
\caption{WMA of the births to teenagers. The figures show the
smooth curves $S_5$, $S_6$, $S_7$, $S_8$. These curves smooth all
fluctuations with a period shorter than 64, 128, 256 and 512 days,
respectively. }
\end{figure}

\newpage

\begin{figure}[tbp]
\epsfig{file=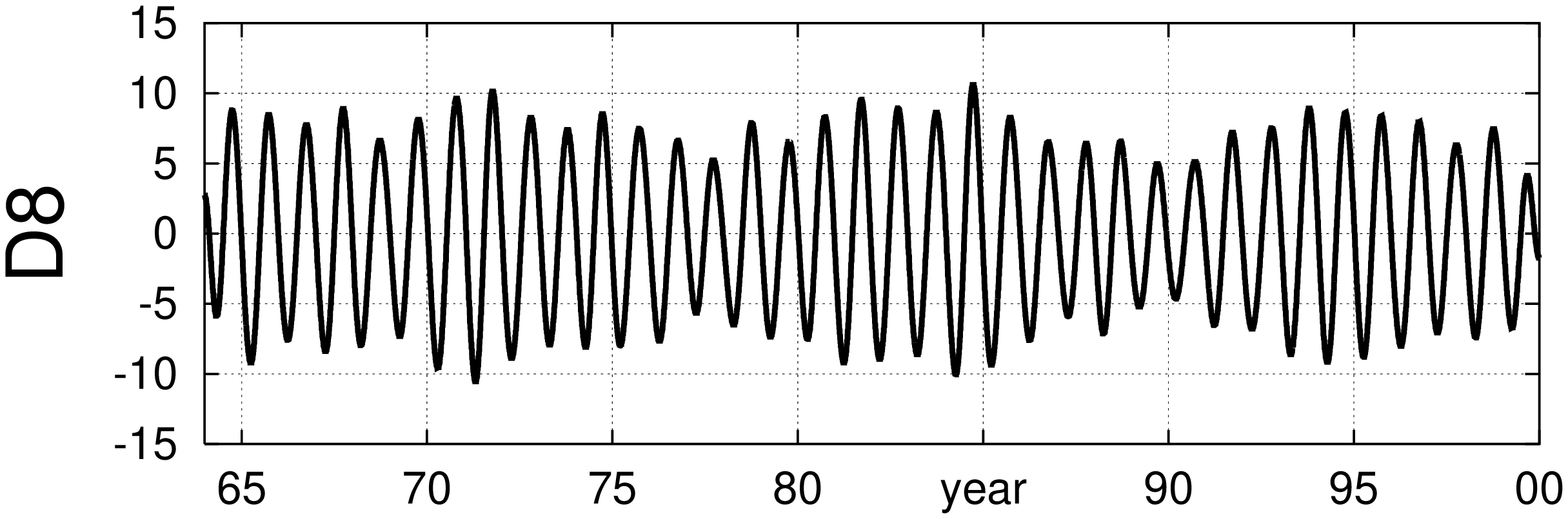,height=15cm,width=4.5cm,angle=-90}

\epsfig{file=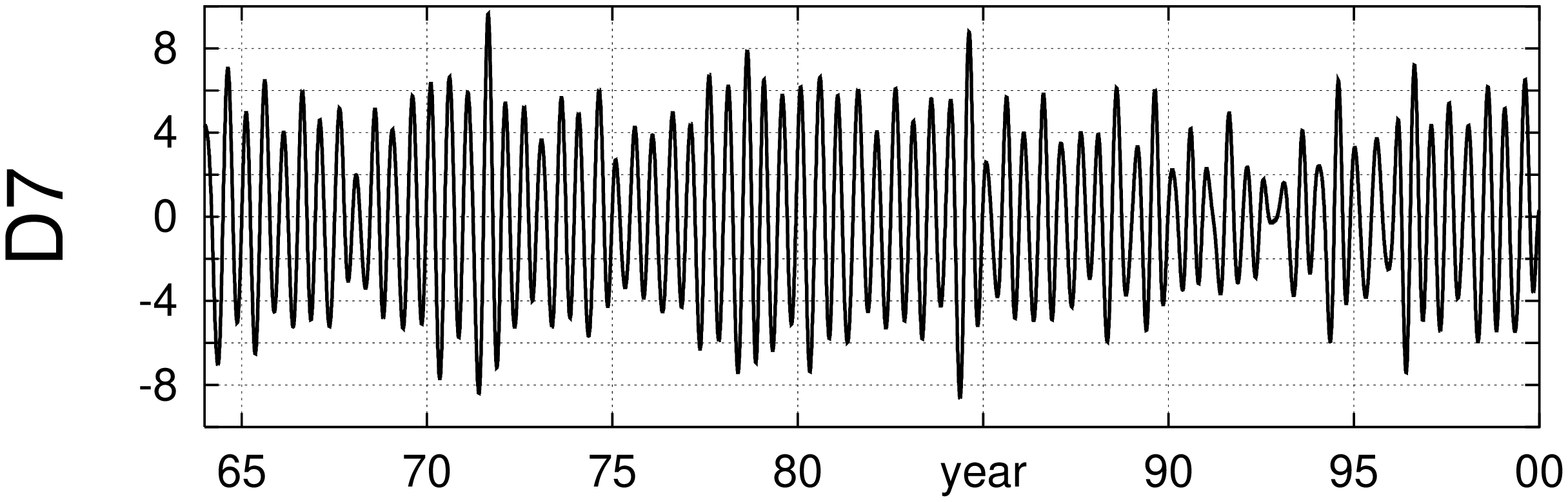,height=15cm,width=4.5cm,angle=-90}

\epsfig{file=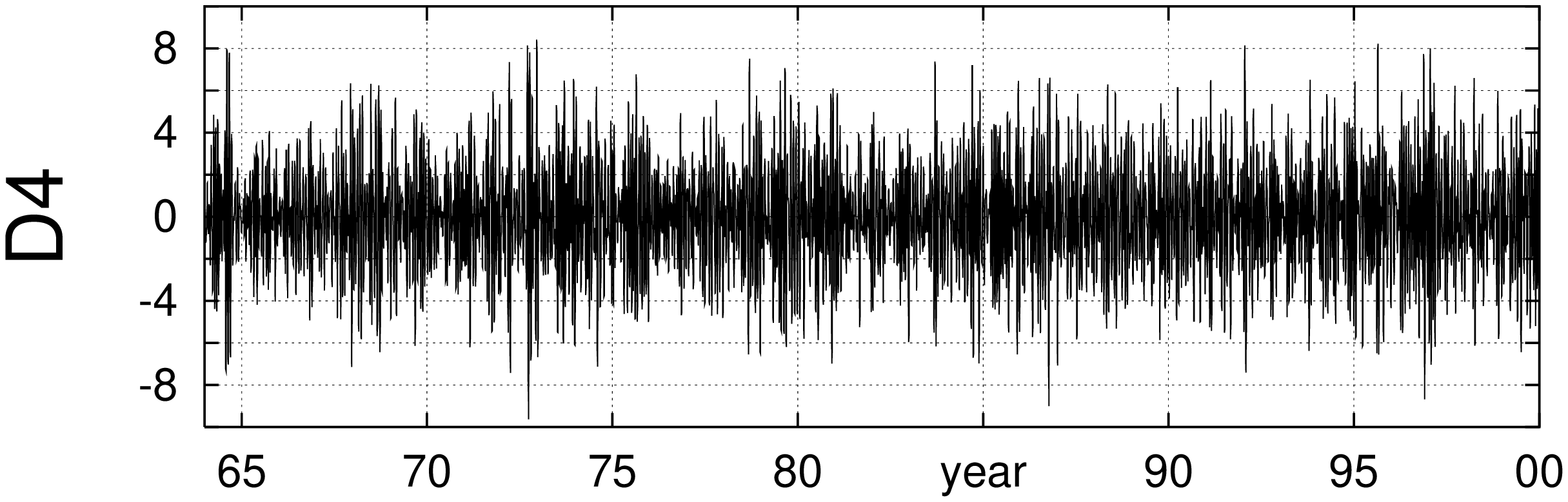,height=15cm,width=4.5cm,angle=-90}

\epsfig{file=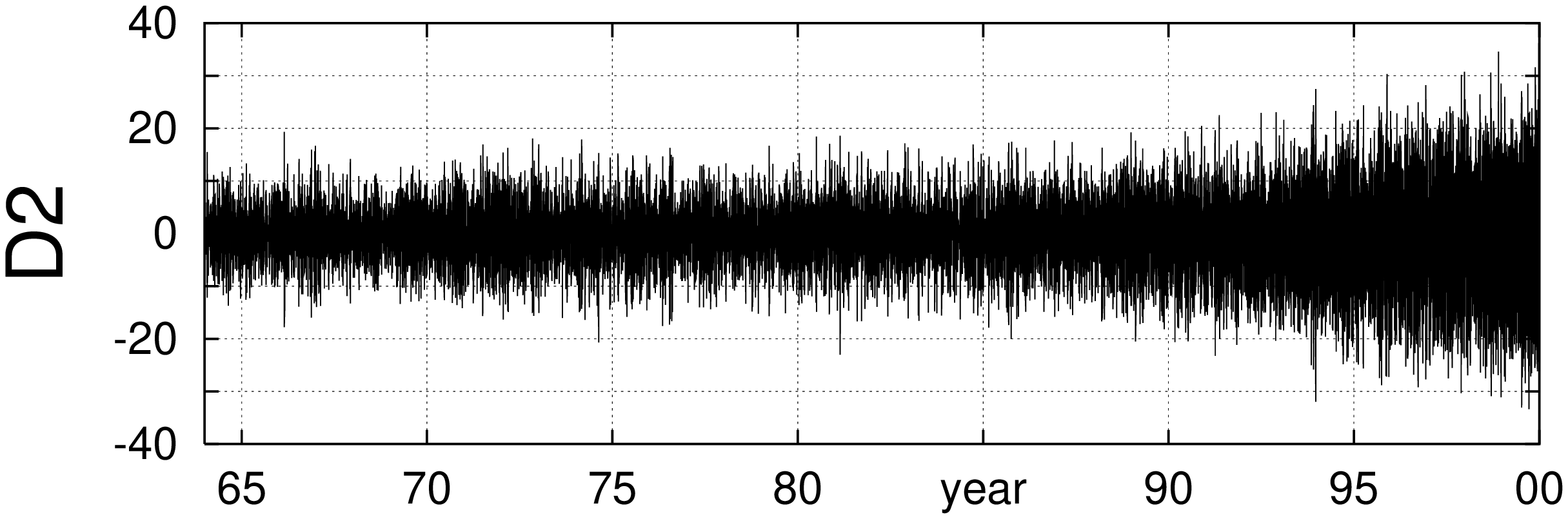,height=15cm,width=4.5cm,angle=-90}
\caption{WMA of the births to teenagers. The figures show the
detail curves $D_2$, $D_4$, $D_7$ and $D_8$. These curves contain
fluctuations with a period within the range of [4:8], [16:32],
[128:256] and [256:512] days, respectively.}
\end{figure}

\newpage

\begin{figure}[tbp]
\epsfig{file=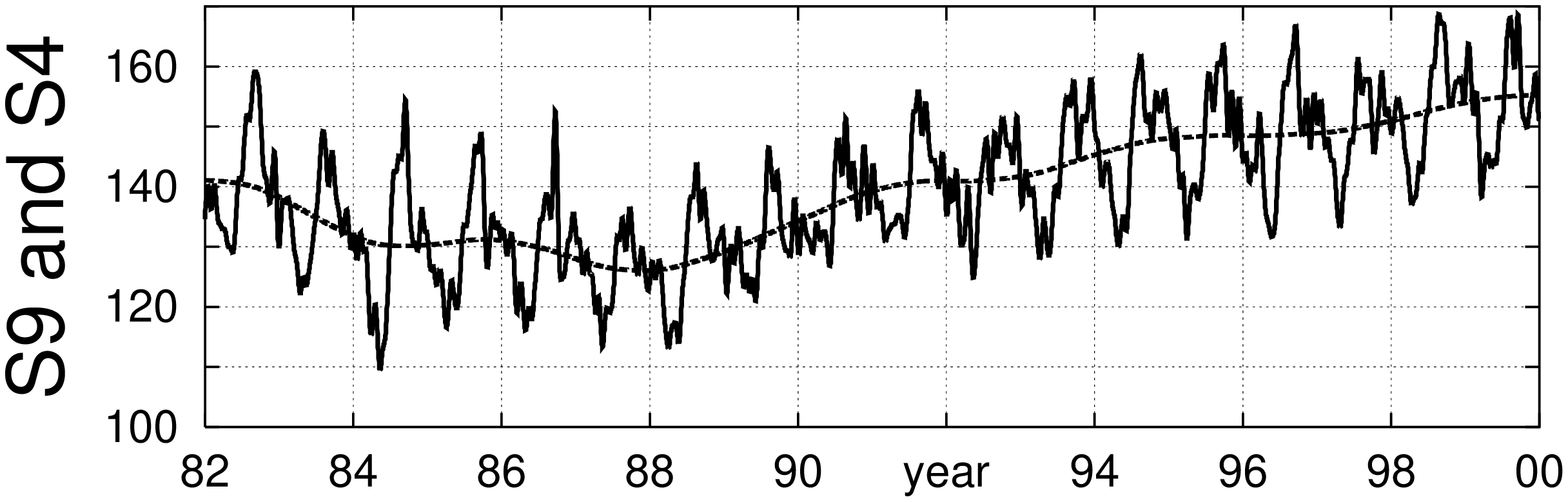,height=15cm,width=4.5cm,angle=-90}

\epsfig{file=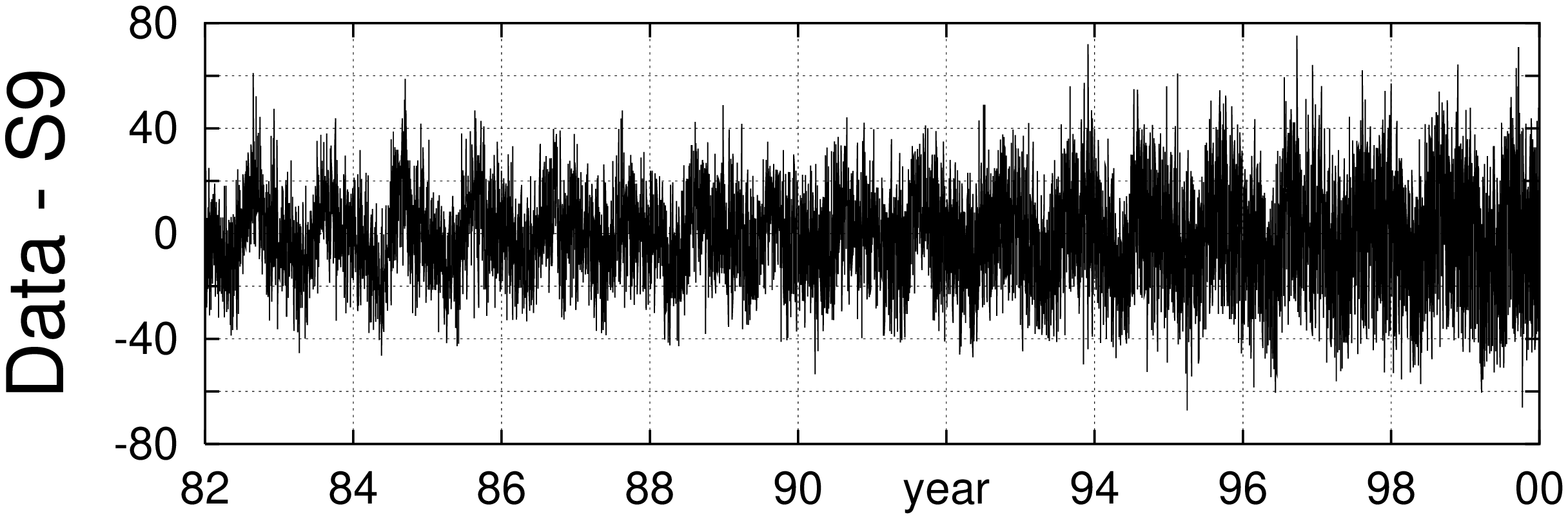,height=15cm,width=4.5cm,angle=-90}

\epsfig{file=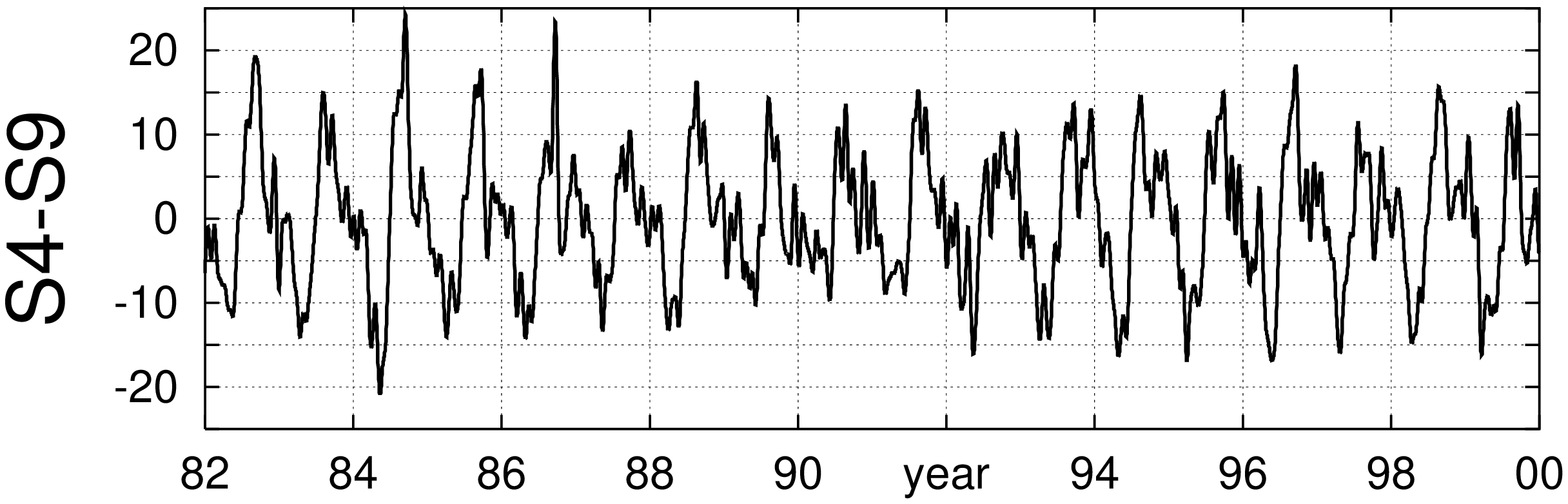,height=15cm,width=4.5cm,angle=-90}
\caption{WMA of the births to teenagers. [a] The first picture
shows the smooth curves $S_4$ (full line) and $S_9$ (dashed line)
that smooth all fluctuations with period shorter than 32 and 1024
days, respectively. [b] The second picture shows the original data
with the smooth curve $S_9$ detrended from them. [c] The third
figure shows the smooth curve $S_4$, with the smooth curve $S_9$
detrended from it.}
\end{figure}

\newpage

\begin{figure}[tbp]
\epsfig{file=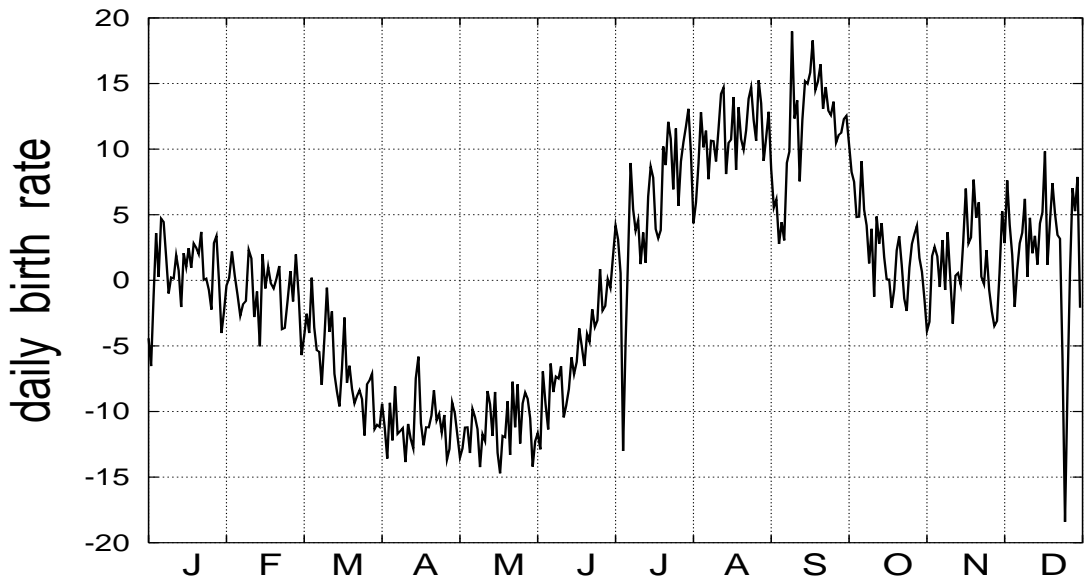,height=15cm,width=8cm,angle=-90}
\caption{ Daily birth rate through the months of the year. These values are obtained from
the 36 yearly cycles of Fig. 6, obtained by detrending the smooth curve $S_9$
from the original data. For each day of the year we evaluate the mean value
of the birth rate over the 36 years of the 1964-1999 time period. The lowest
rates occur in April and May, and the highest in August and September. There
are deep spikes in the correspondence of holidays: first days of January,
4th of July, first week of September and 25th-26th of December.}
\end{figure}

\newpage

\begin{figure}[tbp]
\epsfig{file=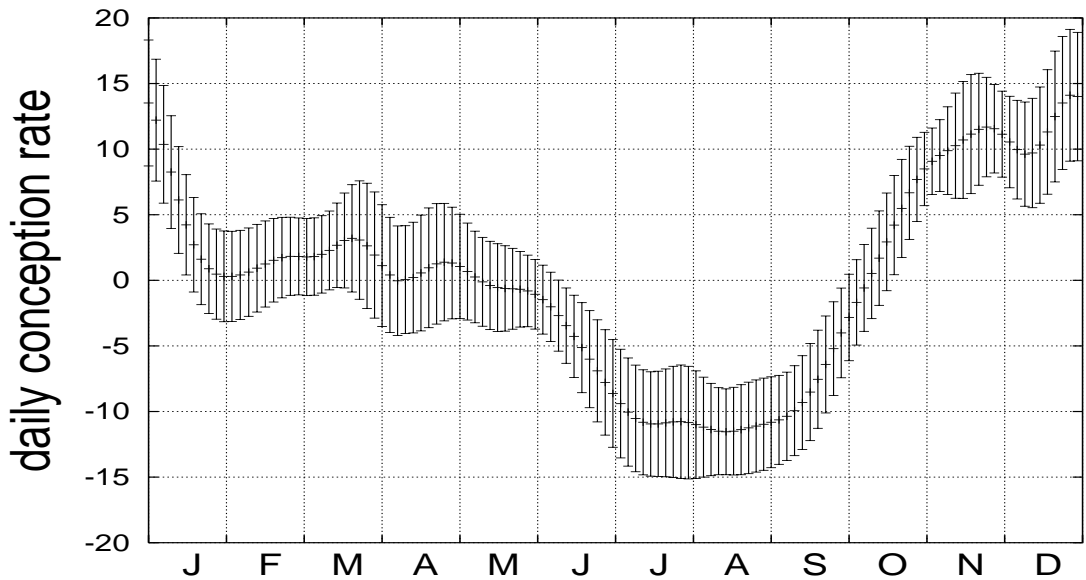,height=15cm,width=8cm,angle=-90}
\caption{Daily conception rate through the months of the year. The values are obtained from the smooth
curve $S_4$, with the smooth curve $S_9$ detrended from it (Fig. 7). For each day
of the year we evaluate the mean value over the 36 years of the 1964-1999
time period. The conception day is estimated to occur 38 weeks before the
delivery. The lowest rate is in July and August, the highest in November and
December. There is a conception rate increase during the holidays: Spring
break in March, Thanksgiving in November and Christmas in December. A fast
growth of the conception rate occurs in September and October.}
\end{figure}

\newpage

\begin{figure}[tbp]
\epsfig{file=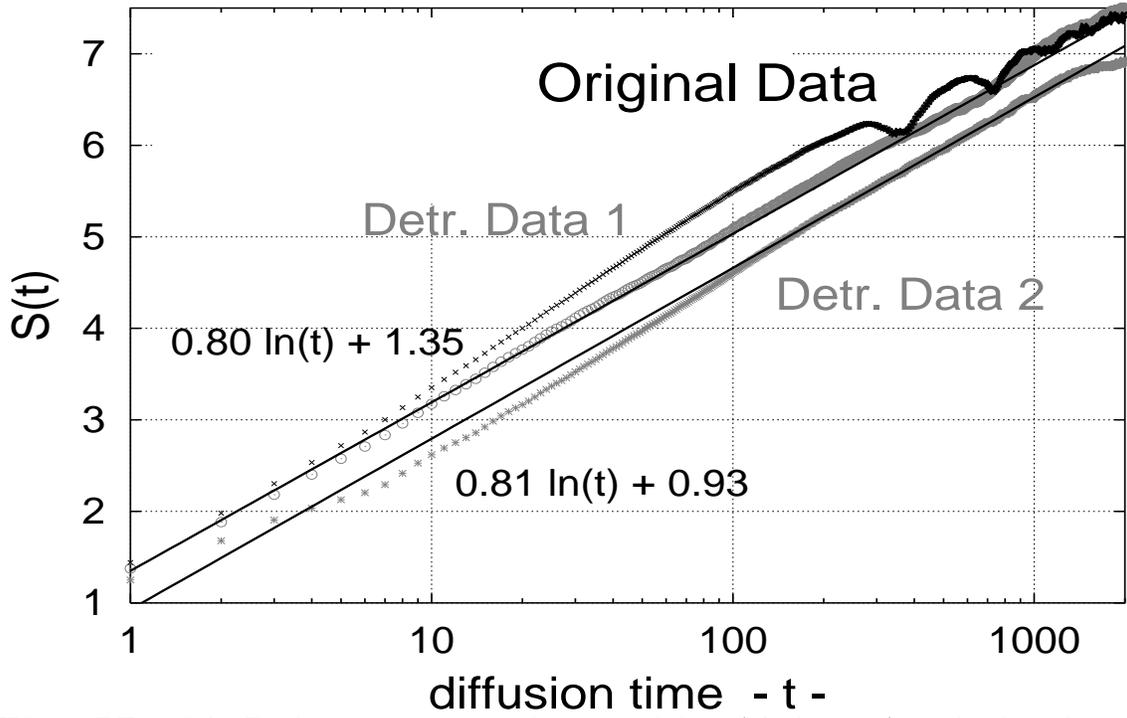,height=16cm,width=10cm,angle=-90}
\caption{DEA of the Births to teenagers of the original data (black curve),
of the data detrended of the details $D_8$ and $D_7$ that correspond to the annual
and one half annual periodicity (gray circle curve) and of the data
detrended (dotted line) of the details $D_8$ and $D_7$ plus the linear fit $y =
a+b(t-1964)$ discussed in Sec. 2 (gray star curve). The slope of the
straight line, $\delta=0.80 \pm 0.02$, is the scaling coefficient of the
background noise according to Eq. (\ref{scafun14}). }
\end{figure}

\newpage

\begin{figure}[tbp]
\epsfig{file=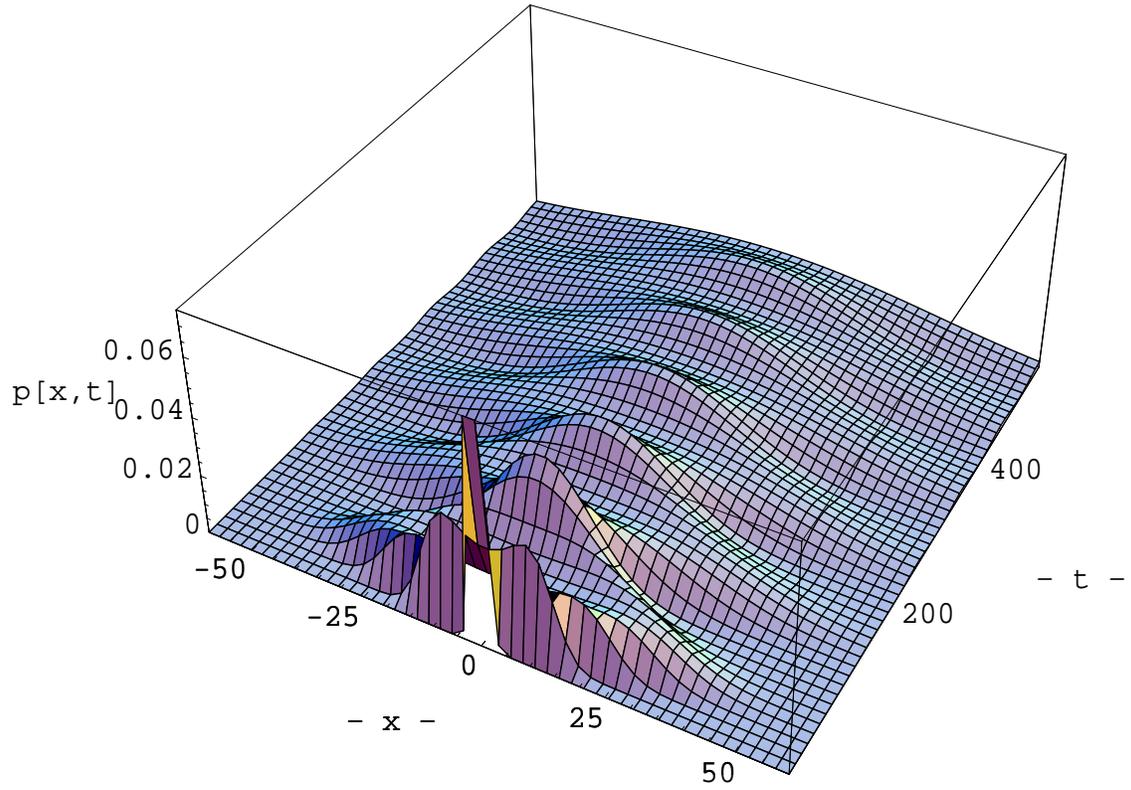,height=13cm,width=15cm,angle=0}
\caption{The figure shows the pdf $p(x,t)$ given by Eq. (\ref{finpdf}). This
pdf is the convolution of the Gaussian diffusion pdf $p_{s}(x,t)$ of Eq. (%
\ref{gaussdistr}) and of the deterministic pdf $p_{d}(x,t)$ of Eq. (\ref
{pdfd}). The period of the deterministic component is $T=100$ and the
amplitude is $A=1$. By increasing the diffusion time $t$ the effect of the
deterministic component becomes weaker and weaker and the Gaussian component
prevails in accordance to the Central Limit Theorem.}
\end{figure}

\newpage

\begin{figure}[tbp]
\epsfig{file=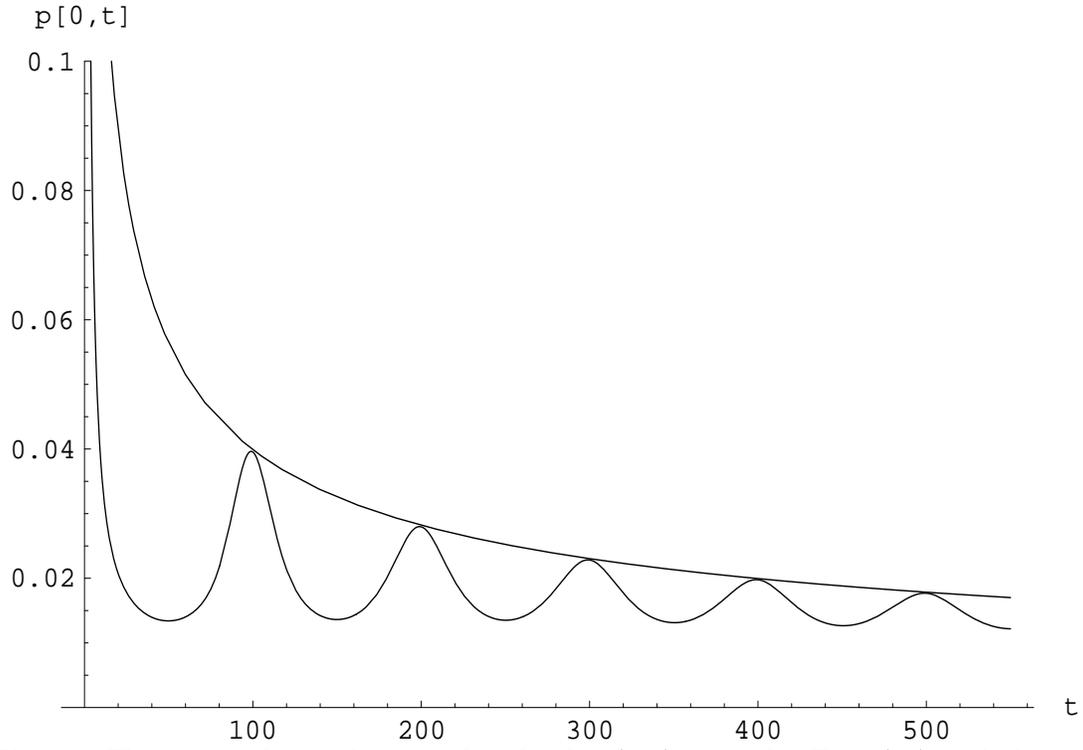,height=10cm,width=15cm,angle=0}
\caption{The figure shows the convoluted pdf $p(x,t)$ given by Eq. (\ref
{finpdf}) and the Gaussian pdf $p_{s}(x,t)$ of Eq. (\ref{gaussdistr}) (upper
curve) at $x=0$. The period of the deterministic component is $T=100$ and
the amplitude is $A=1$. By increasing the diffusion time $t$ the two curves
converge in accordance to the Central Limit Theorem. }
\end{figure}

\newpage

\begin{figure}[tbp]
\epsfig{file=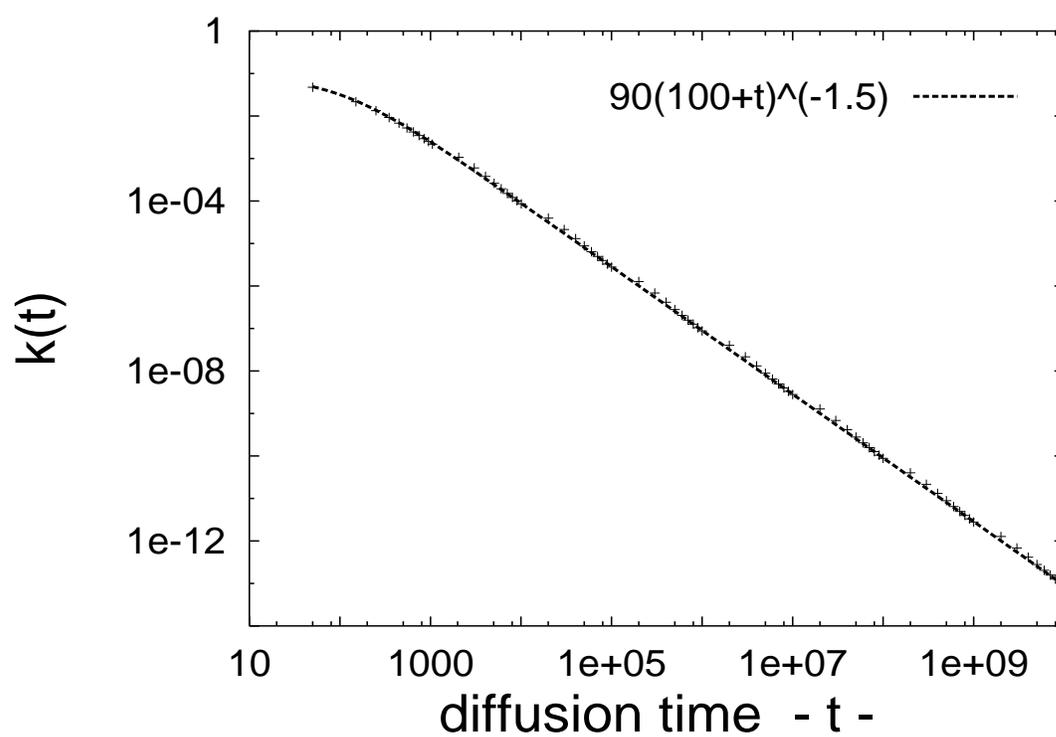,height=15cm,width=10cm,angle=-90}
\caption{Difference $k(t)$ against the diffusion time $t$. According to the
theory, $k(t)\simeq t^{-1.5}$ for large $t$. }
\end{figure}

\newpage

\begin{figure}[tbp]
\epsfig{file=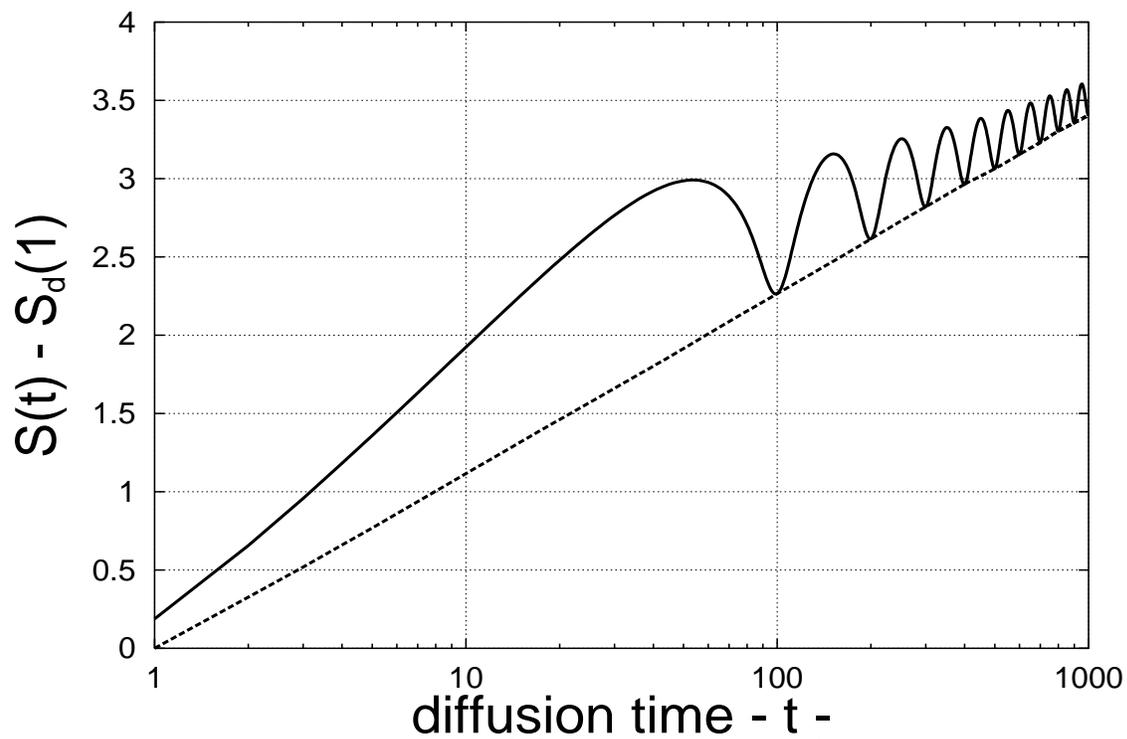,height=15cm,width=10cm,angle=-90}
\caption{DEA of the artificial dataset $\{\zeta_i\}$, Eq. (\ref{newseq}),
(solid line) and of the Gaussian artificial dataset $\{\eta_i\}$ (dashed
line). The dashed curve coincides with a straight line with the slope $%
\delta=0.5$. The constant $S_d(1)$ is the value of the entropy of the
artificial dataset $\{\eta_i\}$ at the first step of diffusion. }
\end{figure}

\newpage

\begin{figure}[tbp]
\epsfig{file=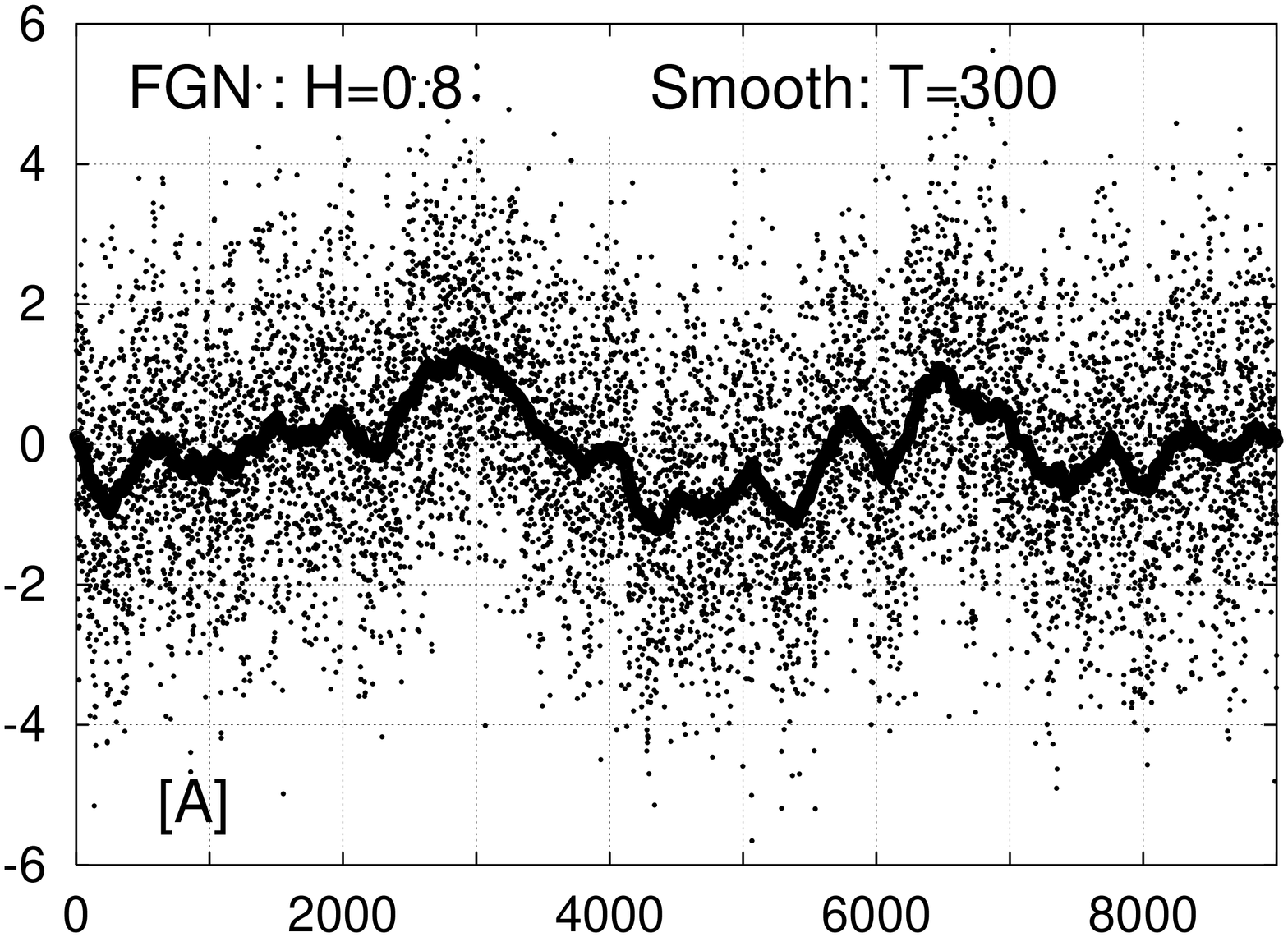,height=7.5cm,width=6cm,angle=-90}
\epsfig{file=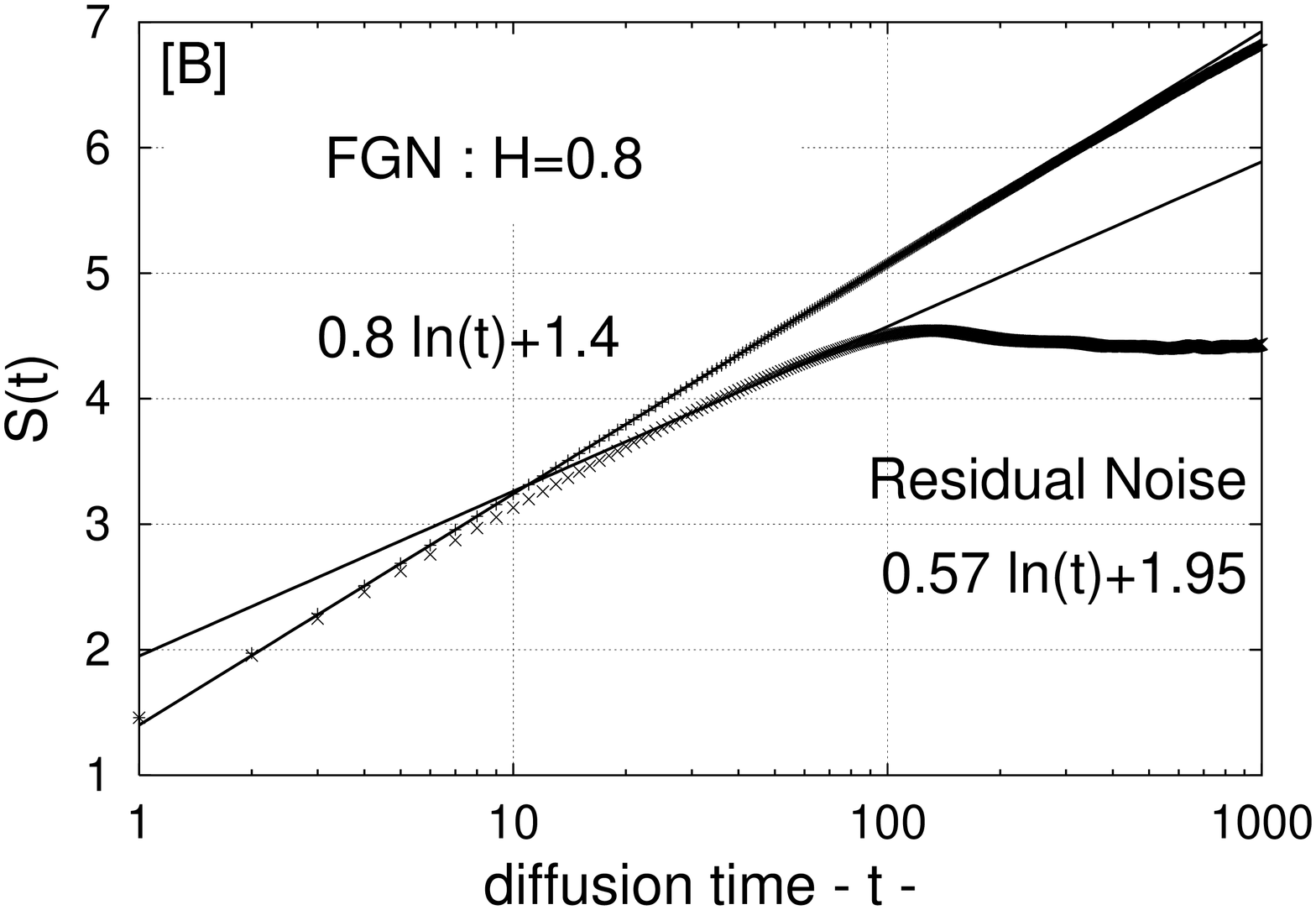,height=7.5cm,width=6cm,angle=-90}
\caption{[a] Time series of an artificial fractional Gaussian
noise with H=0.80 and its smooth curve obtained with a moving
average of period $T= 300$ (generic units). [b] Diffusion entropy
analysis of the original fractal noise (upper curve) and of the
residual noise (lower curve) obtained by detrending the original
noise of the smooth component. The figure shows that the smoothing
detrending procedure disrupts completely the scaling and fractal
properties of the original signal.}
\end{figure}

\newpage

\begin{figure}[tbp]
\epsfig{file=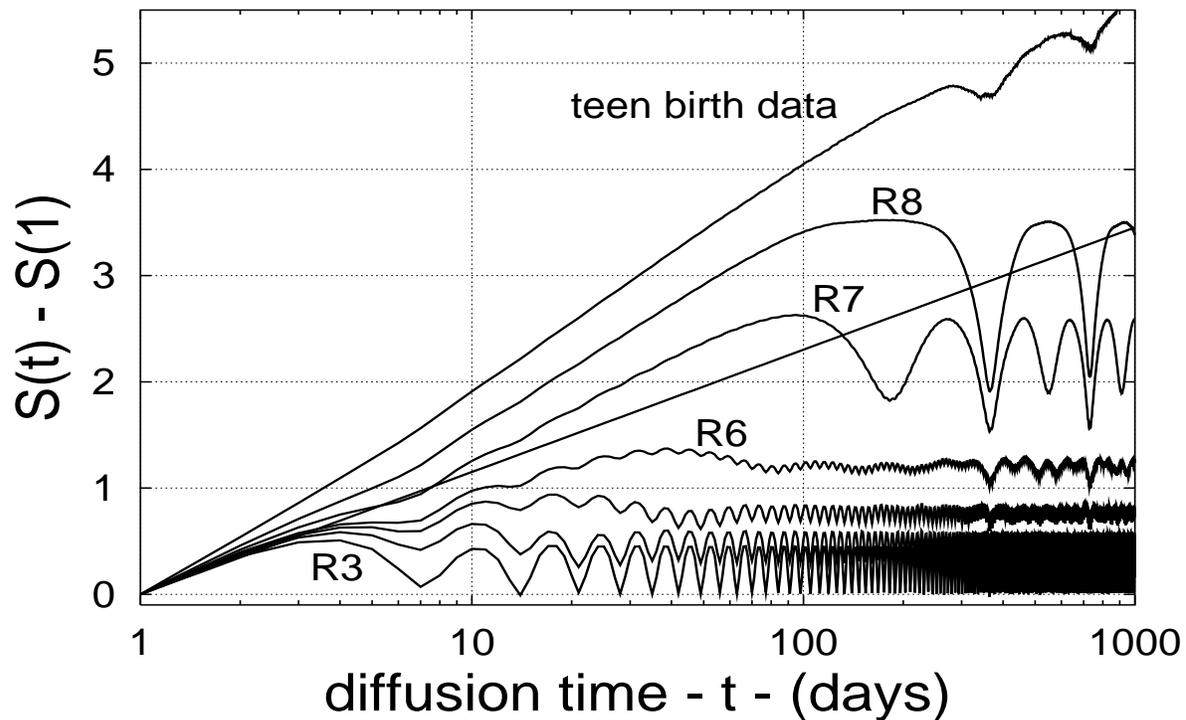,height=16cm,width=10cm,angle=-90}\\

\caption{MDEA of the Births to teenagers. From top to down, the curves are
the DEA of the original data, and of the residuals $R_8$, ..., $R_3$. The straight
line is $f_{DE}(t)=0.5 \ln(t)$ that corresponds to the Gaussian diffusion.
Above the straight line there is the persistent region and below the
straight line there is the antipersistent region.}
\end{figure}

\begin{table}[tbp]
\begin{tabular}{|c|c|c|c|c|}
& S(1) & S$_{sat}$ & S$_{sat}$ & period \\
&  & rel-val & abs-val & days \\ \hline
data & 1.45 & Min : Max & Min : Max &  \\ \hline
$R_8$ & 1.22 & 2.00 : 3.50 & 3.22 : 4.72 & 365 \\ \hline
$R_7$ & 1.19 & 1.60 : 2.60 & 2.79 : 3.79 & 183 \\ \hline
$R_6$ & 1.18 & 1.00 : 1.30 & 2.18 : 2.48 & 75 \\ \hline
$R_5$ & 1.17 & 0.65 : 0.85 & 1.82 : 2.02 & 7 \& 45 \\ \hline
$R_4$ & 1.15 & 0.30 : 0.60 & 1.45 : 1.75 & 7 \& 21 \\ \hline
$R_3$ & 1.03 & 0.00 : 0.44 & 1.03 : 1.47 & 7
\end{tabular}
\caption{ Summary of the information contained in Fig. 14. The first column
reports the values of the entropy at the first step of diffusion, S(1), for
the original data and their rests $R_j$ for $j=2,...,8$. The second and
third columns report the relative and the absolute values of the range of
the height of the horizontal lines that measure the information or entropy
that corresponds to each wavelet scale. The absolute values of the entropy
of the third column are obtained by summing the values of column 1 and 2.
The last column reports the main periodicity present in each rest $R_j$. The
rests $R_8$ and $R_7$ are dominated by the annual periodicity and its harmonic.
The rests $R_5$, $R_4$, $R_3$ are dominated by the strong weekly periodicity. The
values are compatible with those given by the spectral analysis of Fig. 2. }
\end{table}


\end{document}